\def\Statusstring{Submitted to IEEE Transactions on Information Theory \\
                  February 25, 2006}

\documentclass[11pt,letter]{article}

\usepackage{amsmath,amscd,fancyhdr,graphics,pifont}
\usepackage{floatflt,subfigure,epsfig,moreverb,multicol,lscape}
\usepackage{supertabular,tabularx,multirow,bar,amssymb}
\usepackage{theorem,psfrag,afterpage,curves,epic, bbm}
\usepackage[english]{babel}

\newcommand{\ignore}[1]{}

\newcommand{\supp}{\operatorname{supp}}

\newcommand{\dint}[1]{\operatorname{d}{#1}}

\newcommand{\dmins}{d_{\mathrm{min}}}
\newcommand{\matr}[1]{\mathbf{#1}}
\newcommand{\vect}[1]{\mathbf{#1}}
\newcommand{\code}[1]{\mathcal{#1}}
\newcommand{\set}[1]{\mathcal{#1}}
\newcommand{\graph}[1]{\mathsf{#1}}

\newcommand{\GF}[1]{\mathbb{F}_{#1}}
\newcommand{\R}{\mathbb{R}}
\newcommand{\Rp}{\mathbb{R}_{+}}
\newcommand{\Rpp}{\mathbb{R}_{++}}

\newcommand{\tr}{\mathsf{T}}

\newcommand{\PG}[2]{\operatorname{PG}(#1,#2)}
\newcommand{\EG}[2]{\operatorname{EG}(#1,#2)}
\newcommand{\EGq}{\operatorname{EG}(2,q)}
\newcommand{\PGq}{\operatorname{PG}(2,q)}
\newcommand{\codePG}[2]{\code{C}_{\PG{#1}{#2}}}
\newcommand{\codeEG}[2]{\code{C}_{\EG{#1}{#2}}}
\newcommand{\codePGq}{\code{C}_{\PGq}}
\newcommand{\codeEGq}{\code{C}_{\EGq}}
\newcommand{\wcol}{w_{\mathrm{col}}}
\newcommand{\wrow}{w_{\mathrm{row}}}
\newcommand{\defeq}{\triangleq}
\newcommand{\setDML}[1]{\set{D}^{\mathrm{ML}}_{#1}}
\newcommand{\setDMLzero}{\set{D}^{\mathrm{ML}}_{\vect{0}}}
\newcommand{\setDLPzero}{\set{D}^{\mathrm{LP}}_{\vect{0}}}
\newcommand{\vlambda}{\boldsymbol{\lambda}}
\newcommand{\vomega}{\boldsymbol{\omega}}

\newcommand{\convhull}{\operatorname{conv}}

\newcommand{\onenorm}[1]{\lVert #1 \rVert_1}
\newcommand{\twonorm}[1]{\lVert #1 \rVert_2}

\newcommand{\vc}{\vect{c}}
\newcommand{\vs}{\vect{s}}

\newcommand{\vx}{\vect{x}}
\newcommand{\vy}{\vect{y}}
\newcommand{\Z}{\mathbb{Z}}

\newcommand{\hvx}{\vect{\hat x}}

\newcommand{\setJodd}{\set{J}_{\mathrm{odd}}}

\renewcommand{\leq}{\leqslant}
\renewcommand{\geq}{\geqslant}

\newtheorem{Lemma}{Lemma}
\newtheorem{Theorem}[Lemma]{Theorem}

\newtheorem{Corollary}[Lemma]{Corollary}

\theoremstyle{plain}

\newtheorem{PreDefinition}[Lemma]{{\textbf{Definition}}}
  \newenvironment{Definition}%
    {\begin{PreDefinition}}{\hfill$\square$\end{PreDefinition}}

\theoremstyle{plain}

\newtheorem{PreRemark}[Lemma]{{\textbf{Remark}}}
  \newenvironment{Remark}%
    {\begin{PreRemark}\upshape}{\hfill$\square$\end{PreRemark}}

\newtheorem{PreExample}[Lemma]{{\textbf{Example}}}
  \newenvironment{Example}%
    {\begin{PreExample}\upshape}{\hfill$\square$\end{PreExample}}

\newenvironment{Proof}%
  {\noindent \emph{Proof:}}{\hfill$\square$}

\newtheorem{Conjecture}[Lemma]{Conjecture}

\newcommand{\elemma}{\hfill$\square$}

\newcommand{\ecorollary}{\hfill$\square$}

\newcommand{\RM}{\operatorname{RM}}

\newcommand{\rank}{\operatorname{rank}}
\newcommand{\fp}[1]{\set{#1}}
\newcommand{\fph}[2]{\set{#1}(\matr{#2})}
\newcommand{\fc}[1]{\set{#1}}
\newcommand{\fch}[2]{\set{#1}(\matr{#2})}
\newcommand{\Mps}{\set{M}_{\mathrm{p}}}

\newcommand{\wps}{w_{\mathrm{p}}}
\newcommand{\wpsAWGNC}{w_{\mathrm{p}}^{\mathrm{AWGNC}}}
\newcommand{\wpsBSC}{w_{\mathrm{p}}^{\mathrm{BSC}}}
\newcommand{\wpsBEC}{w_{\mathrm{p}}^{\mathrm{BEC}}}

\newcommand{\wpsAWGNCmin}{w_{\mathrm{p}}^{\mathrm{AWGNC,min}}}
\newcommand{\wpsBSCmin}{w_{\mathrm{p}}^{\mathrm{BSC,min}}}
\newcommand{\wpsBECmin}{w_{\mathrm{p}}^{\mathrm{BEC,min}}}

\newcommand{\card}[1]{\lvert #1 \rvert}

\newcommand{\wH}{w_{\mathrm{H}}}
\newcommand{\wHmin}{w_{\mathrm{H}}^{\mathrm{min}}}

\newcommand{\chicw}[1]{\chi_{#1}^{\mathrm{cw}}}
\newcommand{\chimcw}[1]{\chi_{#1}^{\mathrm{mcw}}}
\newcommand{\chiAWGNCmpcw}[1]{\chi_{#1}^{\mathrm{mpcw}, \mathrm{AWGNC}}}
\newcommand{\chiBSCmpcw}[1]{\chi_{#1}^{\mathrm{mpcw}, \mathrm{BSC}}}
\newcommand{\chiBECmpcw}[1]{\chi_{#1}^{\mathrm{mpcw}, \mathrm{BEC}}}
\newcommand{\gAWGNC}[1]{g_{#1}^{\mathrm{AWGNC}}}
\newcommand{\gBSC}[1]{g_{#1}^{\mathrm{BSC}}}
\newcommand{\gBEC}[1]{g_{#1}^{\mathrm{BEC}}}

\newcommand{\type}{t}
\newcommand{\vtype}{\vect{t}}

\newcommand{\setL}{\set{L}}
\newcommand{\setLAWGNC}{\set{L}_{\mathrm{AWGNC}}}
\newcommand{\setLBSC}{\setL_{\mathrm{BSC}}}
\newcommand{\setLBEC}{\setL_{\mathrm{BEC}}}

\begin{document}

\title{Pseudo-Codeword Analysis of \\
       Tanner Graphs
       from Projective and Euclidean Planes%
       \footnote{The first author was partially supported by NSF Grant
                 ITR-0205310.  The second author was supported by NSF Grants
                 ATM-0296033 and DOE SciDAC and by ONR Grant
                 N00014-00-1-0966. The material in this paper has been
                 presented in part at the 2005 International Symposium on
                 Information Theory, Adelaide, Australia, Sep.~4--9, 2005, and
                 at the 43rd Annual Allerton Conference on Communications,
                 Control, and Computing, Monticello, IL, USA, Sep.~28--30.}}

\author{Roxana Smarandache%
  \thanks{Department of Mathematics and Statistics,
          San Diego State University, 
          San Diego, CA 92182, USA. 
          On leave at Department of Mathematics,
          University of Notre Dame,
          Notre Dame, IN 46556, USA.
          Email: \texttt{rsmarand@nd.edu}.
          }
  \ and Pascal O.~Vontobel%
  \thanks{Department of EECS, Massachusetts Institute of Technology, 77
          Massachusetts Avenue, Cambridge, MA 02139, USA. 
          Part of the work was done
          while being with the ECE Department, University of
          Wisconsin-Madison, 1415 Engineering Drive, Madison, WI 53706, USA.
          Email: \texttt{pascal.vontobel@ieee.org}. P.O.V.~is the 
          corresponding author.}
  }

\date{}

\maketitle

\vspace{-6cm}
\begin{flushright}
  \texttt{\Statusstring}\\[1cm]
\end{flushright}
\vspace{+4cm}

\begin{abstract}
  In order to understand the performance of a code under maximum-likelihood
  (ML) decoding, one studies the codewords, in particular the minimal
  codewords, and their Hamming weights. In the context of linear programming
  (LP) decoding, one's attention needs to be shifted to the pseudo-codewords,
  in particular to the minimal pseudo-codewords, and their pseudo-weights. In
  this paper we investigate some families of codes that have good properties
  under LP decoding, namely certain families of low-density parity-check
  (LDPC) codes that are derived from projective and Euclidean planes: we study
  the structure of their minimal pseudo-codewords and give lower bounds on
  their pseudo-weight.
\end{abstract}

\noindent\textbf{Index terms} --- 
  Linear programming decoding, 
  message-passing iterative decoding,
  minimal codewords,
  minimal pseudo-codewords,
  pseudo-weight, 
  codes from projective planes,
  codes from Euclidean planes,
  pseudo-weight spectra.

\section{Introduction}
\label{sec:introduction:1}

The motivation of this paper to look at minimal codewords and minimal
pseudo-codewords is twofold. On the one hand we would like to give performance
guarantees on linear programming (LP) decoding~\cite{Feldman:03:1,
Feldman:Wainwright:Karger:05:1}, which is a decoding method that has recently
emerged as an interesting approach to decoding codes. On the other hand, the
connection made by Koetter and Vontobel~\cite{Koetter:Vontobel:03:1,
Vontobel:Koetter:05:1:subm, Vontobel:Koetter:04:2} between message-passing
iterative (MPI) decoding and LP decoding suggests that results for LP decoding
have immediate implications for MPI decoding.

Apart from presenting some bounds on pseudo-weights that hold in general, this
paper focuses solely on certain families of codes based on projective and
Euclidean planes. One of the reasons for this choice is that in the past,
several groups of authors have experimentally observed that such codes can
perform very well under MPI decoding, see
e.g.~\cite{Lucas:Fossorier:Kou:Lin:00:1, Kou:Lin:Fossorier:01:1}, and
therefore these families of codes are a worthwhile object of study for making
some first steps towards a rigorous understanding of the observed
behavior. Another reason is that these families of codes have concise
descriptions and large automorphism groups which may potentially be used to
simplify their analysis.

The rest of this paper is structured as follows. In
Sec.~\ref{sec:family:of:codes:1} we will introduce the two main families of
codes under investigation. After having reviewed maximum liklihood (ML)
decoding and LP decoding in Sec.~\ref{sec:ml:lp:decoding:1}, in
Secs.~\ref{sec:minimal:codewords:1}
and~\ref{sec:fc:minimal:pseudo:codewords:1} we will explain why minimal
codewords and minimal pseudo-codewords are important in the understanding of
the performance of ML and LP decoding,
respectively. Sec.~\ref{sec:spectra:examples:1} will present pseudo-weight
spectra of some selected codes based on finite geometries. Then, in
Sec.~\ref{sec:bounds:awgnc:pseudo:weight:1} we will analyze the possible
pseudo-weights of pseudo-codewords and we will give upper and lower bounds on
the pseudo-weight of certain vectors. In
Sec.~\ref{sec:effective:minimal:pseudo:codewords:1} we introduce the concept
of effective minimal pseudo-codewords and in
Sec.~\ref{sec:structure:of:minimal:pseudo-codewords:1} we study the structure
of minimal pseudo-codewords. Finally, in Sec.~\ref{sec:conclusions:1} we
offer some conclusions.

In the following, $\R$, $\Rp$, and $\Rpp$ will be the set of real numbers, the
set of non-negative real numbers, and the set of positive real numbers,
respectively. Moreover, the \emph{support} of a vector $\vect{\vx}$ will be
defined as $\supp(\vx) \defeq \big \{ \big. i \, \big| \, x_i \neq 0 \big\}$,
the \emph{Hamming weight} of a vector will be as usual $\wH(\vx) \defeq
|\supp(\vx)|$, and $\langle \vect{a}, \vect{b} \rangle \defeq
\sum_{i} a_i b_i$ will denote the \emph{standard inner product} of two vectors
of equal length.

\section{The Code Families under Investigation}
\label{sec:family:of:codes:1}

The codes under investigation come from the families of codes that were called
type-I PG-LDPC and type-I EG-LDPC codes
in~\cite{Kou:Lin:Fossorier:01:1}. Type-I PG-LDPC codes are defined as
follows. Let $q \defeq 2^s$ for some positive integer $s$ and consider a
(finite) projective plane $\PGq$ (see e.g.~\cite{Batten:97, Hirschfeld:98:1})
with $q^2 + q + 1$ points and $q^2 + q + 1$ lines: each point lies on $q+1$
lines and each line contains $q+1$ points.\footnote{Note that the ``$2$'' in
$\PGq$ stands for the dimensionality of the geometry, which in the case of
planes is $2$.} A standard way of associating a parity-check matrix $\matr{H}$
of a binary linear code to a finite geometry is to let the set of points
correspond to the columns of $\matr{H}$, to let the set of lines correspond to
the rows of $\matr{H}$, and finally to define the entries of $\matr{H}$
according to the incidence structure of the finite geometry. In this way, we
can associate to the projective plane $\PGq$ the code $\codePGq$ with
parity-check matrix $\matr{H} \defeq \matr{H}_{\PGq}$, whose parameters are:
\begin{center}
  \begin{tabular}{ll}
    length                                         & $n      = q^2 + q + 1$, \\
    dimension                                      & $k      = n - 3^s - 1$, \\
    minimum Hamming distance                       & $\dmins = q + 2$, \\
    uniform column weight of $\matr{H}$            & $\wcol  = q + 1$, \\
    uniform row weight of $\matr{H}$               & $\wrow  = q + 1$, \\
    size of $\matr{H}$                             & $n \times n$.
  \end{tabular}
\end{center}
In the usual way, we associate a Tanner graph
$\graph{T}(\matr{H}_{\PGq})$,~\cite{Tanner:81}, to the parity-check matrix
$\matr{H}_{\PGq}$: this graph consists of $n = q^2 + q + 1$ variable nodes of
degree $\wcol = q + 1$ and of $n = q^2 + q + 1$ check nodes of degree $\wrow =
q + 1$.

Type-I EG-LDPC codes are defined as follows. Let $q \defeq 2^s$ for some
positive integer $s$ and consider a (finite) Euclidean plane $\EGq$ (see
e.g.~\cite{Batten:97, Hirschfeld:98:1}) with $q^2$ points and $q^2 + q$ lines:
each point lies on $q+1$ lines and each line contains $q$ points. We
essentially use the same procedure as outlined above in order to associate a
parity-check matrix to a finite geometry. But before doing this, we modify the
Euclidean plane slightly: we select a point of $\EGq$ and remove it together
with the $q+1$ lines through it. Doing so, we obtain an $\EGq$-based code
$\codeEGq$ with parity-check matrix $\matr{H} \defeq \matr{H}_{\EGq}$, whose
parameters are:
\begin{center}
  \begin{tabular}{ll}
    length                                         & $n      = q^2 - 1$, \\
    dimension                                      & $k      = n - 3^s + 1$, \\
    minimum Hamming distance                       & $\dmins = q + 1$, \\
    uniform column weight of $\matr{H}$            & $\wcol  = q$, \\
    uniform row weight of $\matr{H}$               & $\wrow  = q$, \\
    size of $\matr{H}$                             & $n \times n$.
  \end{tabular}
\end{center}
Again, we can associate a Tanner graph
$\graph{T}(\matr{H}_{\EGq})$,~\cite{Tanner:81}, to the parity-check matrix
$\matr{H}_{\EGq}$: this graph consists of $n = q^2 - 1$ variable nodes of
degree $\wcol = q$ and of $n = q^2 - 1$ check nodes of degree $\wrow =
q$.

Both families of codes have the nice property that, with an appropriate
ordering of the columns and rows, the parity-check matrices are circulant
matrices, meaning that $\codePGq$ and $\codeEGq$ are cyclic codes. This fact
can e.g.~be used for efficient encoding. Such symmetries can also
substantially simplify the analysis. Note that the automorphism groups of
$\codePGq$ and $\codeEGq$ contain many more automorphisms besides the
cyclic-shift-automorphism implied by the cyclicity of the codes.

\section{ML and LP Decoding}
\label{sec:ml:lp:decoding:1}

In this section we briefly review ML and LP decoding~\cite{Feldman:03:1,
Feldman:Wainwright:Karger:05:1}. Consider a binary linear code $\code{C}
\subseteq \{ 0, 1 \}^n$ of length $n$ and dimension $k$ that is used for 
data communication over a memoryless binary-input channel with channel law
$p_{Y|X}(y|x)$. The transmitted codeword will be called $\vx \defeq (x_1,
\ldots, x_n)$ and the received vector will be called $\vy \defeq (y_1,
\ldots, y_n)$. Based on the received vector, we can define the log-likelihood
ratio (LLR) vector $\vlambda \defeq (\lambda_1, \ldots,
\lambda_n) \in (\R \cup \{ \pm \infty \})^n$ to be the vector containing 
the LLRs $\lambda_i \defeq \log \big( p_{Y|X}(y_i|0) / p_{Y|X}(y_i|1) \big)$,
$i = 1, \ldots, n$. Using the canonical embedding of the set $\{ 0, 1 \}$ into
$\R$ and of the set $\code{C}$ into $\R^n$, ML decoding can then be cast as
\begin{align}
  \hvx
    &\defeq 
       \arg \min_{\vx \in \code{C}} \ 
         \langle
           \vx, \vlambda
         \rangle.
             \label{eq:ml:decoder:1}
\end{align}
Letting $\convhull(\code{C})$ be the convex hull of $\code{C}$ in $\R^n$, the
above ML decoding rule can also be formulated as
\begin{align}
  \hvx
    &\defeq
       \arg \min_{\vx \in \convhull(\code{C})}
         \langle
           \vx, \vlambda
         \rangle.
             \label{eq:ml:decoder:2}
\end{align}
Unfortunately, for most codes of interest, the description complexity of
$\convhull(\code{C})$ grows exponentially in the block length and therefore
finding the minimum in \eqref{eq:ml:decoder:2} with a linear programming
solver is highly impractical for reasonably long codes.\footnote{Exceptions to
this observation include for example the class of convolutional codes with not
too many states.}

A standard approach in optimization theory and practice is to replace a
minimization problem by a relaxed minimization problem, in our case we replace
the minimization over $\convhull(\code{C})$ by a minimization over some easily
describable polytope $\fp{P}$ which is a relaxation of $\convhull(\code{C})$:
\begin{align}
  \hvx
    &\defeq \arg \min_{\vx \in \fp{P}} \ 
       \langle
           \vx, \vlambda
         \rangle.
           \label{eq:lp:decoder:1}
\end{align}
If $\fp{P}$ is strictly larger than $\convhull(\code{C})$ then the decision
rule in~\eqref{eq:lp:decoder:1} obviously represents a sub-optimal decoder. A
relaxation that works particularly well for LDPC codes is given by the
following approach~\cite{Feldman:03:1, Feldman:Wainwright:Karger:05:1}. Let
$\code{C}$ be described by an $m \times n$ parity-check matrix $\matr{H}$ with
rows $\vect{h}_1, \vect{h}_2, \ldots, \vect{h}_m$. Then the polytope $\fp{P}
\defeq \fph{P}{H}$, in this context also called the \emph{fundamental
polytope}~\cite{Koetter:Vontobel:03:1, Vontobel:Koetter:05:1:subm}, is defined
as
\begin{align*}
  \fp{P}
    &\defeq
       \bigcap_{j=1}^{m}
         \convhull(\code{C}_j)
  \quad \text{ with } \quad
  \code{C}_j
     \defeq \big\{
              \vx \in \{0, 1\}^n
              \ | \ 
              \langle \vect{h}_j, \vx \rangle = 0
              \, \operatorname{mod}\, 2
            \big\}.
\end{align*}
Note that $\fp{P}$ is a convex set within $[ 0, 1 ]^n$ that contains
$\convhull(\code{C})$, but whose description complexity is typically much
smaller than the one of $\convhull(\code{C})$. Points in the set $\fp{P}$ will
be called \emph{pseudo-codewords}. Because the set $\fp{P}$ is usually
strictly larger than $\convhull(\code{C})$, it can obviously happen that the
decoding rule in~\eqref{eq:lp:decoder:1} delivers a vertex of $\fp{P}$ that is
not a codeword. Such vertices that correspond to pseudo-codewords that are not
codewords are the reason for the sub-optimality of LP decoding
(cf.~\cite{Feldman:03:1, Feldman:Wainwright:Karger:05:1,
Koetter:Vontobel:03:1, Vontobel:Koetter:05:1:subm}).  Note that $\fp{P} =
\fph{P}{H}$ is a function of the parity-check matrix $\matr{H}$ that describes
the code $\code{C}$. Different parity-check matrices for the same code might
therefore lead to different fundamental polytopes. It is worthwhile to remark
though that all these fundamental polytopes have the property that $\fph{P}{H}
\cap \{ 0, 1 \}^n = \code{C}$, i.e.~all points of $\fph{P}{H}$ with integral
coordinates are also codewords~\cite{Feldman:03:1,
Feldman:Wainwright:Karger:05:1}.

\section{Minimal Codewords}
\label{sec:minimal:codewords:1}

In this section we will discuss minimal codewords and explain their importance
with respect to ML decoding. Although ML decoding is often impractical,
knowing bounds on the block error rate of an ML decoder can help in assessing
the performance of sub-optimal but practical decoding algorithms.

\begin{Definition}
  \label{lemma:binary:code:minimal:codeword:properties:1:1}

  Let $\code{C}$ be a linear code. A non-zero codeword $\vx \in \code{C}$ is
  called \emph{minimal} if and only if its support does not (strictly) contain
  the support of any other non-zero codeword as a proper subset. The set of
  all minimal codewords of $\code{C}$ is commonly denoted by
  $\set{M}(\code{C})$.
\end{Definition}

We will henceforth assume that we transmit a binary linear code $\code{C}$
over a binary-input output-symmetric channel
(cf.~e.g.~\cite[Def.~1]{Richardson:Urbanke:01:2}). For this setup, when
studying the ML decoder in \eqref{eq:ml:decoder:1} or \eqref{eq:ml:decoder:2},
we can without loss of generality assume that the zero codeword was sent,
because all decision regions are congruent. The importance of minimal
codewords lies in the following considerations.

\begin{Theorem}[cf.~e.g.~\cite{Agrell:96:1}]
  \label{theorem:minimal:codewords:1}
  
  Let $\code{C}$ be a binary linear code of length $n$ and for $\vx \in
  \code{C}$ let
  \begin{align*}
    \setDML{\vect{x}}
      &\defeq \big\{
                \big. 
                  \vlambda \in \R^n 
                \ \big| \ 
                  \langle \vx', \vlambda \rangle
                    \geq \langle \vx, \vlambda \rangle
                    \text{ for all } \vx' \in \code{C} \setminus \{ \vx \} 
              \big\}
  \end{align*}
  be the region in the LLR space where the ML decoder decides in favor of the
  codeword $\vx$.\footnote{We assume that during ML decoding ties between
  decoding regions are resolved randomly.} Then the decision region
  $\setDML{\vect{x}}$ of a codeword $\vx \in \code{C}$ shares a
  facet\footnote{A facet is an $n-1$ dimensional face of a polytope, see
  e.g.~\cite{Deza:Laurent:97:1}.} with the decision region $\setDMLzero$ of
  the zero codeword if and only if $\vx$ is a minimal codeword.
\end{Theorem}

Therefore, knowing the minimal codewords of the code $\code{C}$ is sufficient
in order to assess its ML decoding performance. In the following, we mention a
few known facts about minimal codewords of codes that helped us in our
investigations in the later chapters of this paper.

\begin{Lemma}[\cite{Hwang:79:1, Ashikhmin:Barg:98:1}]
  \label{lemma:binary:code:minimal:codeword:properties:1}

  Let $\set{M}(\code{C})$ denote the set of all minimal codewords of a binary
  $[n,k,d]$ code $\code{C}$. Then:
  \begin{enumerate}

    \item ${\cal M}(\code{C})$ spans $\code{C}$.

    \item If $\matr{H}(\set{U})$ denotes the submatrix of $\matr{H}$ with
      columns indexed by the set $\set{U} \subseteq \{ 1, 2, \ldots, n\}$, we
      have that a codeword $\vx \in \code{C}$ is minimal if and only if
      $\rank\big( \matr{H}(\supp(\vx)) \big) = |\supp(\vx)| - 1 = \wH(\vx) -
      1$.

    \item If $\vx \in \set{M}(\code{C})$ then $|\supp(\vx)| = \wH(\vx) \leq n
    - k + 1$.

    \item If $\vx \in \code{C}$ and $|\supp(\vect{x})| = \wH(\vx)
      \leq 2 d - 1$ then $\vect{x} \in \set{M}(\code{C})$.

    \item Every non-minimal codeword can be written as a sum of two non-zero
      codewords with disjoint supports. \elemma

  \end{enumerate}
\end{Lemma}

Let us mention some further results about minimal codewords (here we also
include non-binary codes). The problem of classifying the minimal codewords is
completely solved only for $q$-ary Hamming codes, for the second order binary
Reed-Muller codes $\RM(2,m)$, and for the $\Z_4$ Kerdock
codes~\cite{Ashikhmin:Barg:98:1,Ashikhmin:Barg:Cohen:Huguet:95:1,
Massey:93:1}, and partially for the binary Reed-Muller code of order $r$,
$\RM(r,m)$~\cite{Borissov:Manev:Nikova:01:1}. However, in general it seems to
be quite difficult to describe the set of minimal codewords for an arbitrary
linear code even in the binary case.

We remark that in the context of linear codes, Hwang~\cite{Hwang:79:1} was the
first to consider the set of minimal codewords of a code (there called the
``projecting set of a code''). He studied them in connection with two
modifications of the correlation decoding algorithm.\footnote{In the light of
the ML decoder as formulated in~\eqref{eq:ml:decoder:2}, these algorithms can
be seen as variations of the simplex method
(cf.~e.g.~\cite{Bertsimas:Tsitsiklis:97:1}) that minimizes $\langle \vx,
\vlambda \rangle$ over the polytope $\convhull(\code{C})$.}

Minimal codewords and their properties arise also in connection with secret
sharing. As it was first pointed out in~\cite{Massey:93:1}, minimal vectors in
a linear code completely specify the access structure of the linear secret
sharing scheme defined by the code. This line of research was pursued
in~\cite{Ashikhmin:Barg:98:1, Ashikhmin:Barg:Cohen:Huguet:95:1}. We finally
note that minimal vectors were also studied in combinatorics under the concept
of cycles of linear matroids.

\section{The Fundamental Cone, Minimal Pseudo-Codewords, and Spectra}
\label{sec:fc:minimal:pseudo:codewords:1}

In this section we will shift our attention to LP decoding and the objects of
interest: pseudo-codewords and in particular the minimal pseudo-codewords. For
LP decoding of a binary linear code that is used for data transmission over a
binary-input output-symmetric channel, it is sufficient to consider the part
of the fundamental polytope $\fp{P}$ around the vertex $\vect{0}$,
cf.~\cite{Koetter:Vontobel:03:1, Vontobel:Koetter:05:1:subm}, i.e.~the
fundamental cone. (See also~\cite{Feldman:03:1,
Feldman:Wainwright:Karger:05:1} that discuss this so-called
``$\code{C}$-symmetry'' property).

\begin{Lemma}[\cite{Feldman:03:1, Feldman:Wainwright:Karger:05:1, 
                         Vontobel:Koetter:05:1:subm}]
  \label{lemma:fundamental:cone:1}

  Let $\code{C}$ be an arbitrary binary linear code and let $\matr{H}$ be its
  parity-check matrix. We let $\set{J} \defeq \set{J}(\matr{H})$ be the set of
  row indices of $\matr{H}$ and we let $\set{I} \defeq \set{I}(\matr{H})$ be
  the set of column indices of $\matr{H}$, respectively. For each $j \in
  \set{J}$, we let $\set{I}_j \defeq \set{I}_j(\matr{H}) \defeq \big\{ i \in
  \set{I} \ | \ h_{ji} = 1 \big\}$. Let the {\em fundamental cone}
  $\fch{K}{H}$ of $\matr{H}$ be the conic hull of the fundamental polytope
  $\fph{P}{H}$. Then, $\fch{K}{H}$ is the set of vectors $\vomega \in \R^n$
  that satisfy
  \begin{alignat}{2}
    \forall j \in \set{J}, \
      \forall i \in \set{I}_j:
        \quad
    &&
    \sum_{i' \in \set{I}_j \setminus \{ i \}}
      \omega_{i'}
       &
       \geq
        \omega_{i},
          \label{eq:fundamental:cone:ineq:1} \\
    \quad\quad
    \forall i \in \set{I}:
        \quad
    &&
    \omega_i
       &
       \geq 0.
         \label{eq:fundamental:cone:ineq:2}
  \end{alignat}
  \elemma
\end{Lemma}

We note that if $\vomega$ is in $\fch{K}{H}$, then also $\alpha \cdot \vomega$
is in $\fch{K}{H}$ for any $\alpha \in \Rpp$. Moreover, for any $\vomega$ in
$\fch{K}{H}$ there exists an $\alpha \in \Rpp$ (in fact, a whole interval of
$\alpha$'s) such that $\alpha \cdot \vomega$ is in $\fph{P}{H}$.

Let $\setDLPzero \defeq \big\{ \big. \vlambda \in \R^n \ \big|
\ \langle \vomega, \vlambda \rangle \geq 0 \text{ for all } \vomega \in 
\fph{P}{H} \setminus \{ \vect{0} \} \big\}$ be the region where the LP 
decoder decides in favor of the codeword $\vect{0}$.\footnote{We assume that
during LP decoding ties between decoding regions are resolved randomly.} It
can easily been seen that $\setDLPzero = \big\{ \big. \vlambda \in \R^n \
\big| \ \langle \vomega,
\vlambda \rangle \geq 0 \text{ for all } \vomega \in \fch{K}{H}
\big\}$. Therefore, when studying LP decoding it is enough to know
$\fch{K}{H}$; all vectors $\vomega \in
\fch{K}{H}$ will henceforth be called pseudo-codewords. Moreover, two 
pseudo-codewords where one is a positive multiple of the other will be
considered to be equivalent.

A class of pseudo-codewords that will be used a few times later on is the
class of so-called unscaled
pseudo-codewords~\cite{Koetter:Li:Vontobel:Walker:04:1,
Vontobel:Koetter:05:1:subm}. These pseudo-codewords have integer entries and
are derived from codewords in finite covers of the Tanner graph
$\graph{T}(H)$. An important property of an unscaled pseudo-codeword $\vomega$
is that $\vomega \ (\operatorname{mod}\ 2) \in
\code{C}$.\footnote{See~\cite{Koetter:Li:Vontobel:Walker:04:1,
Vontobel:Koetter:05:1:subm} for an exact definition of unscaled
pseudo-codewords and their properties.}

Another important class of pseudo-codewords is the class of so-called minimal
pseudo-codewords:

\begin{Definition}[\cite{Koetter:Vontobel:03:1, Vontobel:Koetter:05:1:subm}]
  \label{def:minimal:pseudo:codewor:1}

  Let $\code{C}$ be an arbitrary binary linear code described by the
  parity-check matrix $\matr{H}$ whose fundamental cone is $\fch{K}{H}$. A
  vector $\vomega \in \fch{K}{H}$ is called a \emph{minimal pseudo-codeword}
  if the set $\{ \alpha \cdot \vomega \ | \ \alpha \in \Rp \}$ is an edge of
  $\fch{K}{H}$. The set of all minimal codewords will be called
  $\Mps(\fch{K}{H})$.\footnote{Note that this definition implies that
  $\vect{0} \notin \Mps(\fch{K}{H})$.}
\end{Definition}

For a given binary linear code $\code{C}$ with parity-check matrix $\matr{H}$,
the importance of the set $\Mps(\fch{K}{H})$ lies in the following fact. From
basic cone properties (cf.~e.g.~\cite{Boyd:Vandenberghe:04:1}), it can easily
be seen that $\setDLPzero = \big\{ \big. \vlambda \in \R^n \ \big| \ \langle
\vomega, \vlambda \rangle \geq 0 \text{ for all } \vomega \in \Mps(\fch{K}{H})
\big\}$. Therefore, the set $\Mps(\fch{K}{H})$ completely characterizes the
behavior of the LP decoder. It can be shown~\cite{Vontobel:Koetter:05:1:subm}
that for any minimal pseudo-codeword $\vomega$ there is an $\alpha \in \Rpp$
such that $\alpha \vomega$ is an unscaled pseudo-codeword, which, among other
things, implies that all components of $\alpha \vomega$ are integers.

Note that the above notion of minimal pseudo-codewords generalizes the notion
of minimal codewords in the following sense: whereas minimal pseudo-codewords
correspond one-to-one to the edges of $\fph{P}{H}$ (or $\fch{K}{H}$) around
$\vect{0}$, the minimal codewords correspond one-to-one to the edges of
$\convhull(\code{C})$ around $\vect{0}$. (Minimal codewords are usually also
minimal pseudo-codewords, but not always, as was remarked
in~\cite{Vontobel:Koetter:05:1:subm}.)

Because of the one-to-one relationship between parity-check matrices and
Tanner graphs, the fundamental cone can also be seen as a function of the
Tanner graph representing a code. Therefore, in order to emphasize the
dependence of minimal pseudo-codewords on the representation of the code, we
will sometimes talk about the minimal pseudo-codewords of a Tanner graph.

The fundamental cone is \emph{independent} of the specific memoryless
binary-input channel through which we are transmitting, however, the influence
of a pseudo-codeword on the LP decoding performance depends very much on what
channel is used. This influence will be measured by a channel-dependent
\emph{pseudo-weight} of pseudo-codewords; these pseudo-weights can be seen as
generalizations of the Hamming weight that has traditionally been used to
assess the performance of a code under ML decoding.

\begin{Definition}
  Let $\vomega \in \Rp^n$. The binary-input additive white Gaussian noise
  channel (AWGNC) pseudo-weight~\cite{Wiberg:96,
  Forney:Koetter:Kschischang:Reznik:01:1, Vontobel:Koetter:05:1:subm} of
  $\vomega$ is defined to be
  \begin{align*}
    \wpsAWGNC(\vomega)
      &\defeq \frac{\onenorm{\vomega}^2}
                   {\twonorm{\vomega}^2},
  \end{align*}
  if $\vomega \neq \vect{0}$ and $\wpsAWGNC(\vomega) \defeq 0$ otherwise, where
  $\onenorm{\vomega}$ and $\twonorm{\vomega}$ are the $\mathcal{L}_1$- and
  $\mathcal{L}_2$-
  norm of $\vomega$, respectively. Let $\vomega' \in \Rp^n$ be a vector with
  the same components as $\vomega$ but in non-increasing order. Introducing
  \begin{align*}
    f(\xi)
    &\defeq \omega'_i \quad (i-1 < \xi \leq i,\ 0 < \xi \leq n), \\
    F(\xi)
    &\defeq \int_{0}^{\xi} f(\xi') \dint{\xi'}, \\
    e &\defeq F^{-1} \left( \frac{F(n)}{2} \right),
  \end{align*}
  the BSC pseudo-weight~\cite{Forney:Koetter:Kschischang:Reznik:01:1,
  Vontobel:Koetter:05:1:subm} is defined to be $\wpsBSC(\vomega) \defeq 2e$ if
  $\vomega \neq \vect{0}$ and $\wpsBSC(\vomega) \defeq 0$ otherwise.  Finally,
  the BEC pseudo-weight~\cite{Forney:Koetter:Kschischang:Reznik:01:1,
  Vontobel:Koetter:05:1:subm} is defined to be
  \begin{align*}
    \wpsBEC(\vomega)
      &= \card{\supp(\vomega)}.
  \end{align*}
\end{Definition}

Note that for $\vx \in \{ 0, 1 \}^n$ we have $\wpsAWGNC(\vx) = \wpsBSC(\vx) =
\wpsBEC(\vx) = \wH(\vx)$. Let us briefly comment on the significance of the 
above pseudo-weights. When transmitting over an AWGNC, it can be shown that
the squared Euclidean distance from the point $+\vect{1}$ in signal space,
which corresponds to the codeword $\vect{0}$, to the plane $\big\{ \vlambda
\in \R^n \ | \ \langle \vomega, \vlambda \rangle = 0 \big\}$ is 
$\wpsAWGNC(\vomega)$. When transmitting over a BSC, the LP decoder decides in
favor of $\vect{0}$ and against $\vomega$ if the number of bit-flips on the
BSC is smaller than $\wpsBSC(\vomega)/2$; on the other hand, there is at least
one pattern with at least $\wpsBSC(\vomega)/2$ bit flips such that the LP
decoder decides in favor of $\vomega$ and against $\vect{0}$, assuming that
ties are resolved randomly. Finally, when transmitting over a BEC, the LP
decoder decides in favor of $\vect{0}$ and against $\vomega$ if the number of
erasures on the BEC is smaller than $\wpsBEC(\vomega)$; on the other hand,
there is a pattern with $\wpsBEC(\vomega)$ erasures such that the LP decoder
decides in favor of $\vomega$ and against $\vect{0}$ (assuming that ties are
resolved randomly). For a more detailed discussion,
see~\cite{Vontobel:Koetter:05:1:subm, Forney:Koetter:Kschischang:Reznik:01:1}.

\begin{Definition}
  \label{def:pseudo:weight:enumerators:1}
\label{def:Hamming:weight:enumerators:1}
   Let $\code{C}$ be an arbitrary binary linear code. We recall the definition
  of the \emph{codeword weight enumerator} 
\begin{align*}
    \chicw{\code{C}}(X)
      &\defeq
         \sum_{\vx \in \code{C}}
           X^{\wH(\vx)},
  \end{align*} and define the \emph{minimal codeword
    weight enumerator}, and the \emph{minimal pseudo-codeword AWGNC
    pseudo-weight enumerator}, to be, respectively, the polynomials (with
  potentially non-integer exponents)
  \begin{align*}
    \chimcw{\code{C}}(X)
      &\defeq
         \sum_{\vx \in \set{M}(\code{C})}
           X^{\wH(\vx)},\\
    \chiAWGNCmpcw{\matr{H}}(X)
      &= \sum_{[\vomega] \in \Mps(\fch{K}{H})}
           X^{\wpsAWGNC(\vomega)}.
  \end{align*}
 The summation in the last enumerator is over all equivalence classes
  of minimal pseudo-codewords.\footnote{Two pseudo-codewords $\vomega,
    \vomega' \in \fch{K}{H}$ are in the same equivalence class if there exists
    an $\alpha \in \Rpp$ such that $\vomega = \alpha \cdot \vomega'$.} (The
  \emph{minimal pseudo-codeword BSC pseudo-weight enumerator} and the
  \emph{minimal pseudo-codeword BEC pseudo-weight enumerator} are defined
  analogously.)
\end{Definition}

Instead of ``weight enumerator'' and ``pseudo-weight enumerator'' we will
frequently use the words ``weight spectrum'' or ``pseudo-weight spectrum'',
respectively, or simply ``spectrum''. Ideally, for a code defined by a
parity-check matrix $\matr{H}$, we would like to know all the terms of the
spectra that were defined in Def.~\ref{def:Hamming:weight:enumerators:1}
. Often, we have to settle with less,
in particular one often focuses on some quantities that characterize important
aspects of a spectrum.

One such quantity is the minimum pseudo-weight: we let
$\wpsAWGNCmin(\matr{H})$, $\wpsBSCmin(\matr{H})$, and $\wpsBECmin(\matr{H})$
be the minimum AWGNC, BSC, and BEC pseudo-weights of a parity-check matrix
$\matr{H}$, i.e.~the minimum of the respective pseudo-weights, over all
non-zero points in $\fch{K}{H}$. Using some simple tree-based techniques, or
Th.~1 in~\cite{Vontobel:Koetter:04:1}, one can show that
$\wpsAWGNCmin(\matr{H}_{\PGq}) \geq q + 2$ and because this lower bound
matches the minimum Hamming weight, we actually know that
$\wpsAWGNCmin(\matr{H}_{\PGq}) = q + 2$. Similarly, one can show that
$\wpsBSCmin(\matr{H}_{\PGq}) = \wpsBECmin(\matr{H}_{\PGq}) = q + 2$ and that
$\wpsAWGNCmin(\matr{H}_{\EGq}) = \wpsBSCmin(\matr{H}_{\EGq}) =
\wpsBECmin(\matr{H}_{\EGq}) = q + 1$.

Another important quantity that characterizes pseudo-weight spectra is the
pseudo-weight spectrum gap:

\begin{Definition}
  \label{def:gap:1}

  Let $\code{C}$ be an arbitrary binary linear code described by the
  parity-check matrix $\matr{H}$ and let $\Mps'(\fch{K}{H})$ be the set of all
  minimal pseudo-codewords that are \emph{not} multiples of minimal codewords.
  We call the real-valued quantity
  \begin{align*}
    \gAWGNC{\matr{H}}
      &\defeq \min_{\vomega \in \Mps'(\fch{K}{H})} \wpsAWGNC(\vomega)
       -
       \wHmin(\code{C}(\matr{H}))
  \end{align*}
  the AWGNC pseudo-weight spectrum gap of $\matr{H}$. (The BSC pseudo-weight
  spectrum gap and the BEC pseudo-weight spectrum gap are defined
  analogously.)
\end{Definition}

Using \cite[Cor.~8]{Koetter:Vontobel:03:1} (see also
\cite[Sec.~7]{Vontobel:Koetter:05:1:subm}), one can show that for a randomly
constructed $(\wcol,\wrow)$-regular code with $3 \leq \wcol < \wrow$ the AWGNC
pseudo-weight spectrum gap becomes \emph{strictly negative} with probability
one as the block length goes to infinity. However, slightly extending the
arguments that lead to the above mentioned minimum pseudo-weight results, one
can show that for the $\PGq$- and $\EGq$-based codes (with square parity-check
matrix as discussed in Sec.~\ref{sec:introduction:1}) the pseudo-weight
spectrum gap is \emph{non-negative} for finite $q$. In fact, we will see that
for the codes investigated in Sec.~\ref{sec:spectra:examples:1} the
pseudo-weight spectrum gap is significantly positive. We note that by applying
simple performance bounding techniques it can be shown that the larger the gap
is, the closer is the LP decoding performance (and potentially also the
iterative decoding performance~\cite{Koetter:Vontobel:03:1,
Vontobel:Koetter:05:1:subm}) to the ML decoding performance as the SNR goes to
infinity.\footnote{Of course, the pseudo-weight spectrum gap is only a first
approximation to how quickly the LP decoding performance approaches the ML
decoding performance as the SNR goes to infinity. A better approximation is
given by initial parts (or the whole) minimal pseudo-codeword pseudo-weight
enumerator.}

\section{Examples of Spectra}
\label{sec:spectra:examples:1}

In this section we exemplify many of the objects that were defined in the
previous chapters, i.e.~we present minimal pseudo-codewords, weight
enumerators, and the pseudo-weight spectrum gap for some short $\PGq$- and
$\EGq$-based codes.

\subsection{Type-I PG-LDPC Code for $q = 2$}
\label{sec:spectra:examples:pg:2:2:1}

The $\PG{2}{2}$-based code $\codePG{2}{2}$ has parameters $[n {=} 7, k {=} 3,
\dmins {=} 4]$ and can be represented by the following circulant parity-check
matrix $\matr{H}_{\PG{2}{2}}$ of size $7 \times 7$:
\begin{align}
  \matr{H}_{\PG{2}{2}}
    &= \begin{pmatrix}
         1 & 1 & 0 & 1 & 0 & 0 & 0 \\
         0 & 1 & 1 & 0 & 1 & 0 & 0 \\
         0 & 0 & 1 & 1 & 0 & 1 & 0 \\
         0 & 0 & 0 & 1 & 1 & 0 & 1 \\
         1 & 0 & 0 & 0 & 1 & 1 & 0 \\
         0 & 1 & 0 & 0 & 0 & 1 & 1 \\
         1 & 0 & 1 & 0 & 0 & 0 & 1
       \end{pmatrix}.
         \label{eq:pg:2:2:pcm:1}
\end{align}
It is not difficult to find out that the set $\set{M}(\code{C}_{\PG{2}{2}})$
of minimal codewords consists of the following codewords:
\begin{align*}
  & (1, 0, 0, 1, 0, 1, 1), \\
  & (1, 1, 0, 0, 1, 0, 1), \\
  & (1, 1, 1, 0, 0, 1, 0), \\
  & (0, 1, 1, 1, 0, 0, 1), \\
  & (1, 0, 1, 1, 1, 0, 0), \\
  & (0, 1, 0, 1, 1, 1, 0), \\
  & (0, 0, 1, 0, 1, 1, 1).
\end{align*}
Obviously, all of them have Hamming weight $4$ and they are all cyclic shifts
of each other. Because the code has $2^3 = 8$ codewords in total, it turns out
that this code is special in the following sense: there are no non-zero
codewords that are not minimal codewords.

The set $\Mps(\fc{K}(\matr{H}_{\PG{2}{2}}))$ of minimal codewords contains all
the elements from $\set{M}(\code{C}_{\PG{2}{2}})$ plus the following
pseudo-codewords that are not codewords (we show one representative per
equivalence class):
\begin{align*}
  & (1, 2, 2, 1, 2, 1, 1), \\
  & (1, 1, 2, 2, 1, 2, 1), \\
  & (1, 1, 1, 2, 2, 1, 2), \\
  & (2, 1, 1, 1, 2, 2, 1), \\
  & (1, 2, 1, 1, 1, 2, 2), \\
  & (2, 1, 2, 1, 1, 1, 2), \\
  & (2, 2, 1, 2, 1, 1, 1).
\end{align*}
All these minimal pseudo-codewords that are not codewords are cyclic shifts of
each other. The weight enumerators are given by:
\begin{align*}
  \chicw{\codePG{2}{2}}(X)
    &= X^0 + 7 X^4, \\
  \chimcw{\codePG{2}{2}}(X)
    &= 7 X^4, \\
  \chiAWGNCmpcw{\matr{H}_{\PG{2}{2}}}(X)
    &= 7 X^4 + 7 X^{6.25}, \\
  \chiBSCmpcw{\matr{H}_{\PG{2}{2}}}(X)
    &= 7 X^4 + 7 X^{5}, \\
  \chiBECmpcw{\matr{H}_{\PG{2}{2}}}(X)
    &= 7 X^4 + 7 X^{7}.
\end{align*}
Hence, the pseudo-weight spectrum gaps are $\gAWGNC{\matr{H}_{\PG{2}{2}}} =
6.25 - 4 = 2.25$, $\gBSC{\matr{H}_{\PG{2}{2}}} = 5 - 4 =
1$, and $\gBEC{\matr{H}_{\PG{2}{2}}} = 7 - 4 = 3$.

The codes introduced in Sec.~\ref{sec:family:of:codes:1} were based on square
parity-check matrices. However, the code $\PG{2}{2}$ can also be described by
a parity-check matrix of size $4 \times 7$ such as
\begin{align}
  \matr{H}'_{\PG{2}{2}}
    &= \begin{pmatrix}
         1 & 1 & 0 & 1 & 0 & 0 & 0 \\
         0 & 1 & 1 & 0 & 1 & 0 & 0 \\
         0 & 0 & 1 & 1 & 0 & 1 & 0 \\
         0 & 0 & 0 & 1 & 1 & 0 & 1
       \end{pmatrix},
         \label{eq:pg:2:2:modified:pcm:1}
\end{align}
which is the matrix formed by the first four lines of the
$\matr{H}_{\PG{2}{2}}$. Because $\matr{H}'_{\PG{2}{2}}$ contains a subset of
the rows of $\matr{H}_{\PG{2}{2}}$ it is clear that
$\fc{K}(\matr{H}_{\PG{2}{2}}) \subseteq
\fc{K}(\matr{H}'_{\PG{2}{2}})$. Moreover, note that a minimal pseudo-codeword
of $\fc{K}(\matr{H}_{\PG{2}{2}})$ does not need to be minimal pseudo-codeword
of $\fc{K}(\matr{H}'_{\PG{2}{2}})$. Indeed, the set of minimal
pseudo-codewords that are not codewords turns out to be the union of the
following sets (in which we show one representative per equivalence class):
\begin{align*}
  &
    \big\{(3, 2, 1, 1, 1, 0, 0)\big\}, 
    \big\{(0, 1, 2, 1, 1, 3, 0)\big\}, 
    \big\{(0, 1, 1, 1, 2, 0, 3)\big\},
    \big\{(0, 1, 1, 1, 1, 0, 0)\big\}, \\
  & \big\{(2, 1, 1, 1, 0, 0, 1),\ 
    (1, 2, 1, 1, 1, 0, 0),\    
    (0, 1, 2, 1, 1, 1, 0)\big\},\\
  &  \big\{  (0, 1, 1, 1, 0, 2, 1), \ 
     (1, 0, 1, 1, 1, 0, 2), \
      (2, 1, 0, 1, 1, 1, 0)\big\},\\
  &  
   \big\{  (0, 1, 1, 1, 2, 0, 1)\   
     (1, 0, 1, 1, 1, 2, 0),\
    (0, 1, 0, 1, 1, 1, 2)\big\},    
\end{align*}
where cyclic shifts of the same pseudo-codeword are grouped in the same set.
It is interesting to see that, for $\matr{H}'_{\PG{2}{2}}$, a cyclic shift of
a minimal pseudo-codeword is not necessarily a (minimal) pseudo-codeword, as
it was in the case of the matrix $\matr{H}_{\PG{2}{2}}$.
 
It follows that
\begin{align*}
  \chiAWGNCmpcw{\matr{H}'_{\PG{2}{2}}}(X)
    &= 11 X^4 + 9 X^{4.5}, \\
  \chiBSCmpcw{\matr{H}'_{\PG{2}{2}}}(X)
    &= 3 X^3 + 17 X^4, \\
  \chiBECmpcw{\matr{H}'_{\PG{2}{2}}}(X)
    &= 8 X^4 + 12 X^5, 
\end{align*}
and that the pseudo-weight spectrum gaps are $\gAWGNC{\matr{H}'_{\PG{2}{2}}}
= 4 - 4 = 0$, $\gBSC{\matr{H}'_{\PG{2}{2}}} = 3 - 4 = -1$, and
$\gBEC{\matr{H}'_{\PG{2}{2}}} = 4 - 4 = 0$. Comparing the enumerator
$\chiAWGNCmpcw{\matr{H}_{\PG{2}{2}}}(X)$ with the enumerator
$\chiAWGNCmpcw{\matr{H}'_{\PG{2}{2}}}(X)$ it is apparent that the performance
of LP decoding using the second representation will be worse than the
performance of LP decoding using the first representation. Based on iterative
decoder simulations, MacKay and Davey~\cite[Sec.~4]{MacKay:Davey:01:1}
observed a similar performance hierarchy between different representations of
the same code. Note that the code under investigation
in~\cite{MacKay:Davey:01:1} was the $\PG{2}{16}$-based code. Similar
statements can be made for the BSC and the BEC.

Before concluding this subsection, let us comment on the vector $\vomega
\defeq (0, 1, 1, 1, 1, 0, 0)$, which is a minimal pseudo-codeword for
$\matr{H}'_{\PG{2}{2}}$ but not a codeword for $\code{C}_{\PG{2}{2}}$, even
though it has only 0 and 1 components. From our remarks after
Lemma~\ref{lemma:fundamental:cone:1}, it follows that $\vomega$ cannot be an
unscaled pseudo-codeword because $\vomega \text{ (mod $2$)}$ is not a
codeword. However, its equivalent $(0, 2, 2, 2, 2, 0, 0)$ is an unscaled
pseudo-codeword, and it stems from a triple cover. 

Non-codeword pseudo-codewords that contain only zero and one components will
be discussed again in Th.~\ref{theorem:awgnc:pseudo:weight:lower:bound:5:1}.

\subsection{Type-I PG-LDPC Code for $q = 4$}
\label{sec:spectra:examples:pg:2:4:1}

\begin{figure}
  \begin{center}
    \epsfig{file=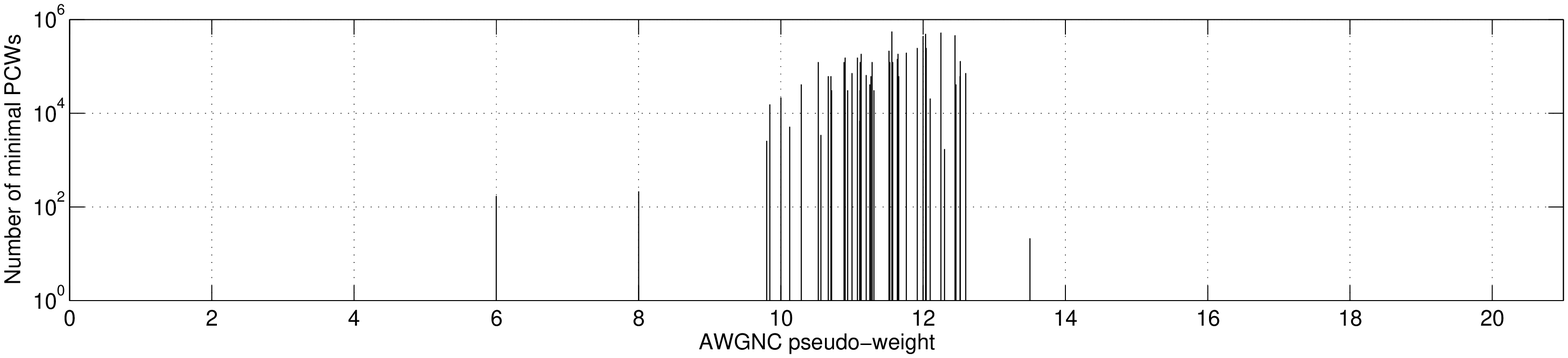,
            width=\linewidth} \\[0.3cm]
    \epsfig{file=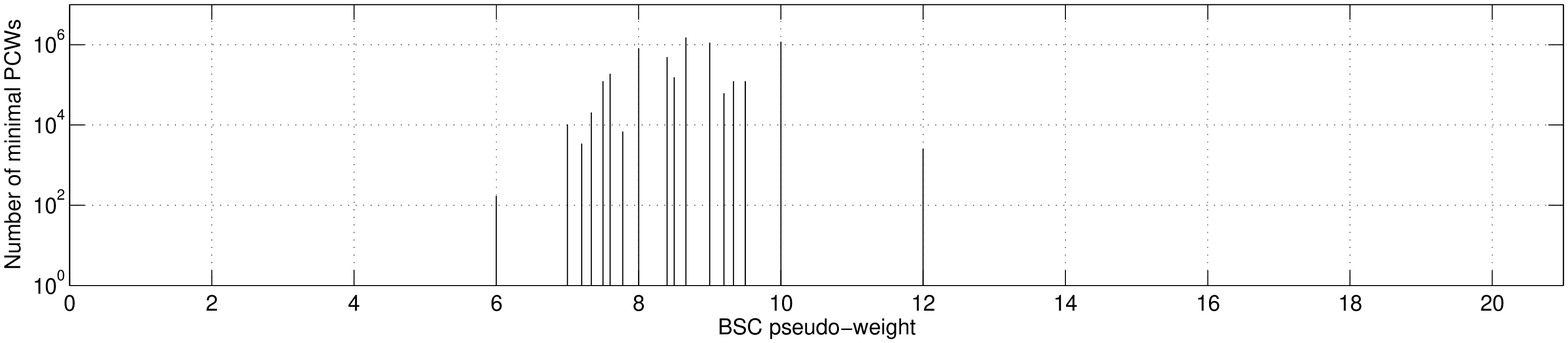,
            width=\linewidth} \\[0.3cm]
    \epsfig{file=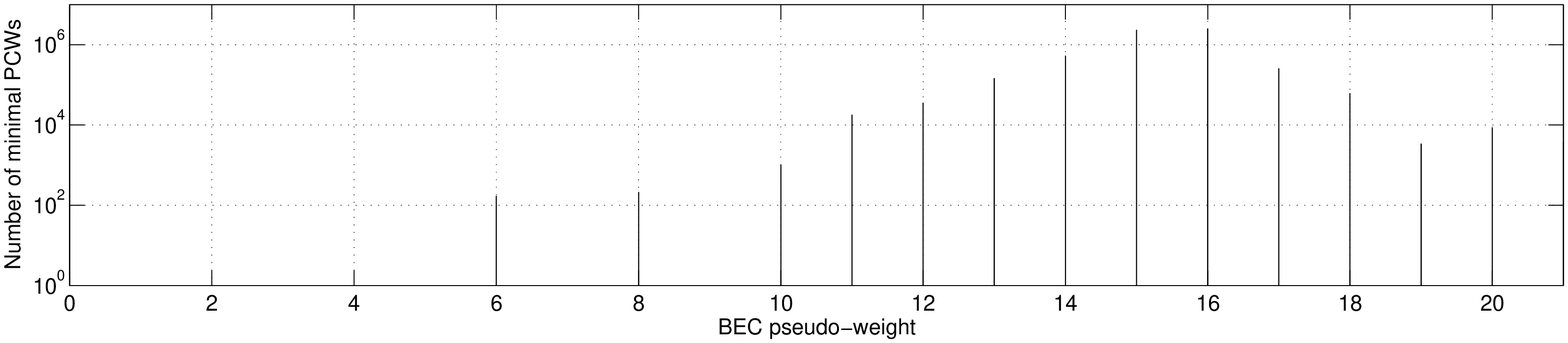,
            width=\linewidth}
  \end{center}
  \caption{Histograms of the AWGNC, BSC, and BEC pseudo-weight of minimal
    pseudo-codewords (PCWs) of the $\PG{2}{4}$-based code. (Note that the
    y-axes are logarithmic.)}
  \label{fig:pg:code:q:4:min:pseudo:codewords:histogram:1}
\end{figure}

The parity-check matrix $\matr{H}_{\PG{2}{4}}$ of the $\PG{2}{4}$-based code
$\codePG{2}{4}$ has size $21 \times 21$, uniform column and row weight $5$,
and yields a code with parameters $[n {=} 21, k {=} 11, \dmins {=} 6]$. The
codeword weight enumerator and the minimal codeword weight enumerator are
\begin{align*}
  \chicw{\codePG{2}{4}}(X)
    &= X^0 + 168 X^6 + 210 X^8 + 1008 X^{10}
       + 280 X^{12} + 360 X^{14} + 21 X^{16}, \\
  \chimcw{\codePG{2}{4}}(X)
    &= 168 X^6 + 210 X^8 + 1008 X^{10},
\end{align*}
respectively. Looking at these enumerators we see that all codewords with
Hamming weight $6$, $8$, and $10$ are minimal codewords. Analyzing the set of
all weight-$6$ codewords one sees that they all have the same pattern,
i.e.~they can all be obtained from a single weight-$6$ codeword by applying a
suitable $\PG{2}{4}$-automorphism. The same is true for all other sets of
codewords with the same weight. This makes the classification of all the
codewords of $\codePG{2}{4}$, and in particular of the minimal codewords of
$\codePG{2}{4}$, relatively easy.

Instead of giving the formula for $\chiAWGNCmpcw{\matr{H}_{\PG{2}{4}}}(X)$,
$\chiBSCmpcw{\matr{H}_{\PG{2}{4}}}(X)$, and
$\chiBECmpcw{\matr{H}_{\PG{2}{4}}}(X)$, we simply show their histogram,
cf.~Fig.~\ref{fig:pg:code:q:4:min:pseudo:codewords:histogram:1}.  Without
going into any details, it is apparent from
Fig.~\ref{fig:pg:code:q:4:min:pseudo:codewords:histogram:1} that the influence
of minimal pseudo-codewords can vary depending on the channel that is
used. (For related observations about varying influences of minimal
pseudo-codewords, see also the discussion in~\cite{Haley:Grant:05:1}.) The
pseudo-weight gaps turn out to be
\begin{alignat*}{2}
  \gAWGNC{\matr{H}_{\PG{2}{4}}}
    &= 9.8 - 6 
   &&= 3.8, \\
  \gBSC{\matr{H}_{\PG{2}{4}}}
    &= 7 - 6 
   &&= 1, \\
  \gBEC{\matr{H}_{\PG{2}{4}}}
    &= 11 - 6
   &&= 5.
\end{alignat*}
We refer to the end of Sec.~\ref{sec:fc:minimal:pseudo:codewords:1} for a
discussion on the significance of positive pseudo-weight gaps.

\subsection{Type-I PG-LDPC Code for $q = 8$}
\label{sec:spectra:examples:pg:2:8:1}

The parity-check matrix $\matr{H}_{\PG{2}{8}}$ of the $\PG{2}{8}$-based code
$\codePG{2}{8}$ has size $73 \times 73$, uniform column and row weight $9$,
and yields a code with parameters $[n {=} 73, k {=} 45, \dmins {=} 10]$.
Judging from some random search experiments in the fundamental cone
$\fc{K}(\matr{H}_{\PG{2}{8}})$, the AWGNC pseudo-weight spectrum gap
$\gAWGNC{\matr{H}_{\PG{2}{8}}}$ seems to be at least~$6.0$.

\subsection{Type-I EG-LDPC Code for $q = 4$}
\label{sec:spectra:examples:eg:2:4:1}

The parity-check matrix $\matr{H}_{\EG{2}{4}}$ of the $\EG{2}{4}$-based code
$\codeEG{2}{4}$ has size $15 \times 15$, uniform column and row weight $4$,
and yields a code with parameters $[n {=} 15, k {=} 7, \dmins {=} 5]$. The
codeword weight enumerator and the minimal codeword weight enumerator are
\begin{align*}
  \chicw{\codeEG{2}{4}}(X)
    &=  X^0 + 18 X^5 + 30 X^6 + 15 X^7
        + 15 X^8 + 30 X^9 + 18 X^{10} + X^{15}, \\
  \chimcw{\codeEG{2}{4}}(X)
    &=  18 X^5 + 30 X^6 + 15 X^7 + 15 X^8 + 30 X^9,
\end{align*}
respectively. Looking at these enumerators we see that all codewords with
Hamming weight $5$, $6$, $7$, $8$, and $9$ are minimal codewords. Analyzing
the set of all weight-$5$ codewords one sees that they all have the same
pattern, i.e.~they can all be obtained from a single weight-$5$ codeword by
applying a suitable $\EG{2}{4}$-automorphism. The same is true for all other
sets of codewords with the same weight.

\begin{figure}
  \begin{center}
    \epsfig{file=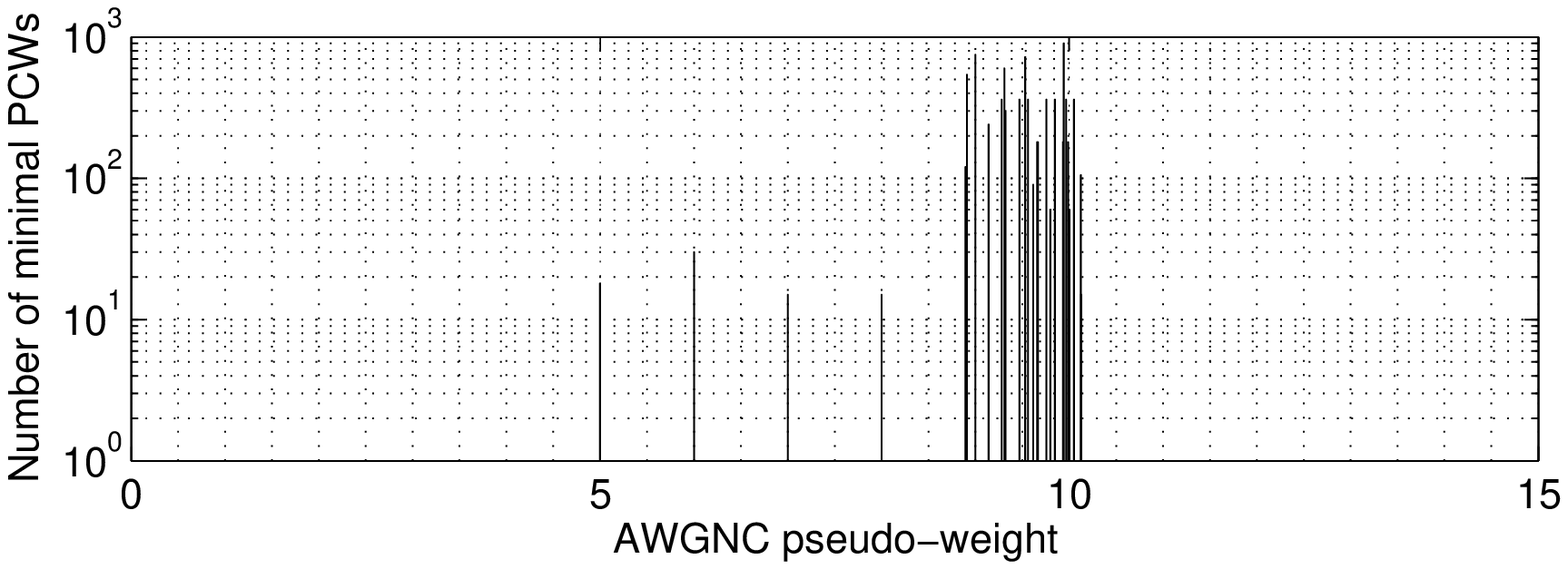,
            width=\linewidth}
    \epsfig{file=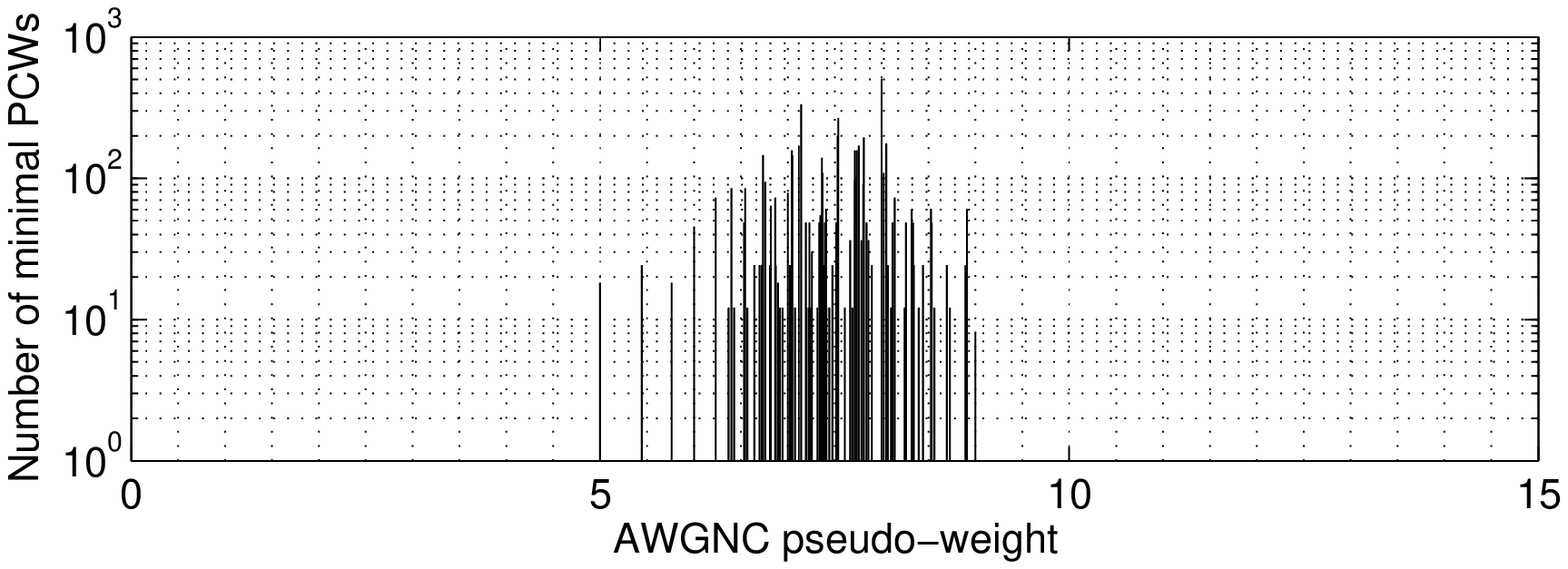,
            width=\linewidth}
    \epsfig{file=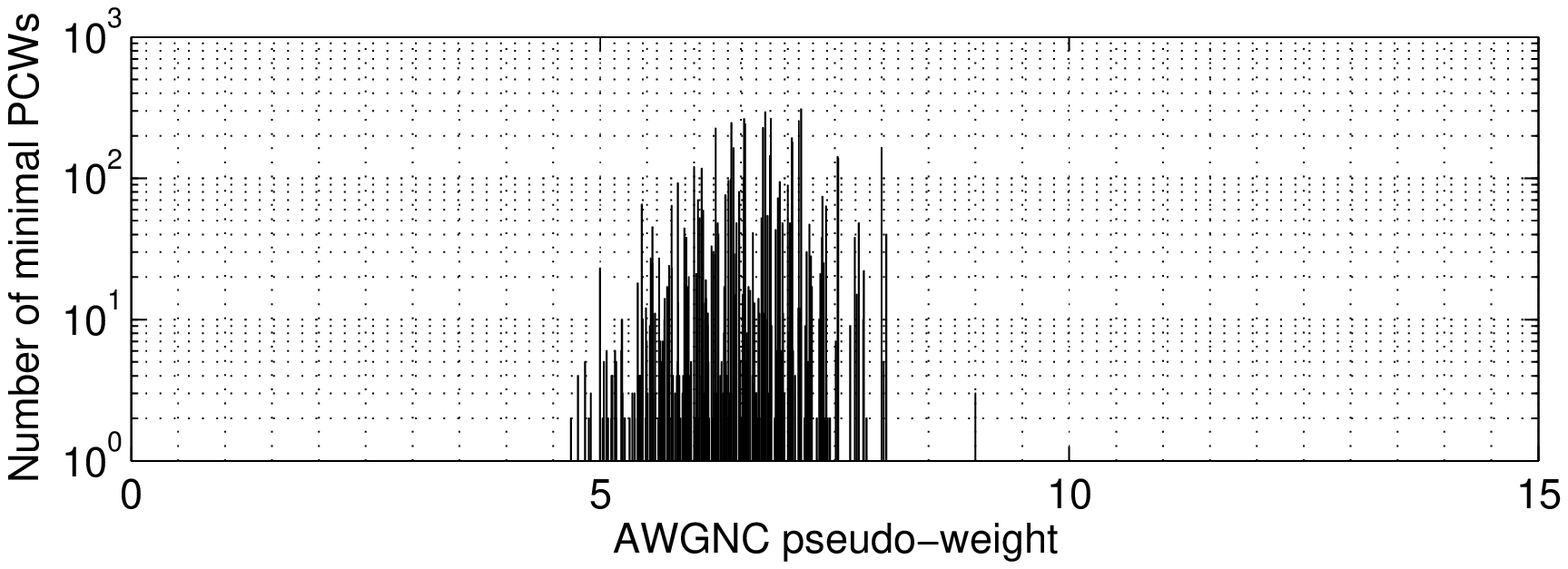,
            width=\linewidth}
  \end{center}
  \caption{Histogram of the AWGNC pseudo-weight of minimal pseudo-codewords
    (PCWs) of the $\EG{2}{4}$-based code. (Note that the y-axes are
    logarithmic.) Top: For $15 \times 15$ parity-check matrix
    $\matr{H}_{\EG{2}{4}}$, $\gAWGNC{\matr{H}_{\EG{2}{4}}} = \frac{169}{19} -
    5 \approx 8.89 - 5 = 3.89$. Middle: For $9 \times 15$ parity-check matrix
    $\matr{H}'_{\EG{2}{4}}$, $\gAWGNC{\matr{H}'_{\EG{2}{4}}} = \frac{49}{9} -
    5 \approx 5.44 - 5 = 0.44$. Bottom: For $8 \times 15$ parity-check matrix
    $\matr{H}''_{\EG{2}{4}}$, $\gAWGNC{\matr{H}''_{\EG{2}{4}}} =
    \frac{361}{77} - 5 \approx 4.69 - 5 = -0.31$.}
  \label{fig:eg:code:q:4:min:pseudo:codewords:histogram:1}
\end{figure}

\begin{figure}
  \begin{center}
    \epsfig{file=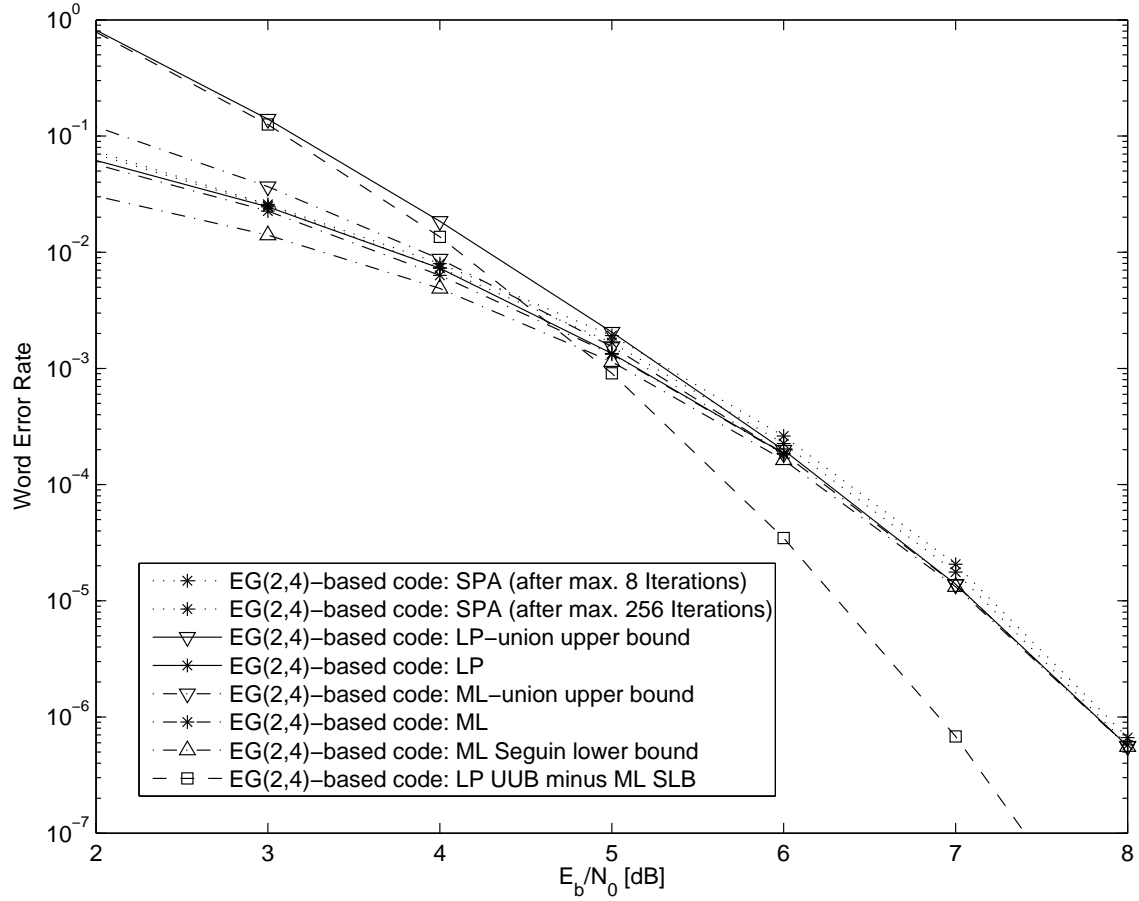,
            width=\linewidth}
  \end{center}

  \caption{Word error rate for various decoding algorithms together with some
    upper and lower bounds. (See main text for explanations.)}
  \label{fig:eg:code:q:4:decoding:and:bounds:3}
\end{figure}

The histograms in Fig.~\ref{fig:eg:code:q:4:min:pseudo:codewords:histogram:1}
correspond to various parity-check matrices that describe $\codeEG{2}{4}$.
Fig.~\ref{fig:eg:code:q:4:min:pseudo:codewords:histogram:1} (top) shows the
histogram for $\chiAWGNCmpcw{\matr{H}_{\PG{2}{4}}}(X)$;
Fig.~\ref{fig:eg:code:q:4:min:pseudo:codewords:histogram:1} (middle) shows the
histogram for $\chiAWGNCmpcw{\matr{H}'_{\PG{2}{4}}}(X)$ where
$\matr{H}'_{\PG{2}{4}}$ is a randomly selected $9 \times 15$ submatrix (with
column weights at least $2$) of $\matr{H}_{\PG{2}{4}}$; and finally
Fig.~\ref{fig:eg:code:q:4:min:pseudo:codewords:histogram:1} (bottom) shows the
histogram for $\chiAWGNCmpcw{\matr{H}''_{\PG{2}{4}}}(X)$ where
$\matr{H}''_{\PG{2}{4}}$ is an $8 \times 15$ submatrix (with five columns
having weight only one) of consecutive rows of the (circulant) matrix
$\matr{H}_{\PG{2}{4}}$. It can easily be seen that for the parity-check
matrices under investigation those with more dependent rows lead to more
favorable histograms.

In Fig.~\ref{fig:eg:code:q:4:decoding:and:bounds:3} we show various decoding
simulation results for data transmission over a binary-input AWGNC and lower
and upper bounds: $\matr{H}_{\EG{2}{4}}$-based sum-product algorithm decoding,
$\matr{H}_{\EG{2}{4}}$-based LP decoding, $\codeEG{2}{4}$-based ML decoding,
an upper bound on LP decoding based on a union of events upper bound, an upper
bound on ML decoding based on a union of events upper bound, and a lower bound
on ML decoding based on an inequality by de Caen as presented by
S\'eguin~\cite{Seguin:98:1}. It can be seen that thanks to the knowledge of
minimal codewords and minimal pseudo-codewords we are able to obtain bounds
that are very tight from a certain SNR value on. This is witnessed by the fast
decreasing line labeled ``LP UUB minus ML SLB'' which shows the difference
between the union upper bound on the LP decoder word error rate and the
S\'eguin lower bound on the ML decoder word error rate.

\subsection{How the Results Were Obtained}
\label{sec:spectra:examples:how:1}

Let us briefly mention how the results for the minimal pseudo-codewords were
obtained in
Secs.~\ref{sec:spectra:examples:pg:2:2:1},~\ref{sec:spectra:examples:pg:2:4:1},
and~\ref{sec:spectra:examples:eg:2:4:1}. We used the program
``lrs''~\cite{Avis:00:1} to search edges in cones. For the code
$\codePG{2}{4}$ in Sec.~\ref{sec:spectra:examples:pg:2:4:1} we additionally
used the two-transitivity of the points of a projective plane in order to
formulate a simpler edge-enumeration subproblem which can be solved
efficiently and from which all the minimal pseudo-codewords can be
derived. This goes as follows: it is clear that a minimal pseudo-codeword
cannot only fulfill inequalities of type~\eqref{eq:fundamental:cone:ineq:2}
with equality (because only $\vomega =
\vect{0}$ does this), i.e.~at least one inequality of
type~\eqref{eq:fundamental:cone:ineq:1} must be fulfilled with
equality. Because the automorphism group of $\PG{2}{4}$ is two-transitive
(which means that for any two pair of points there exists an automorphism that
maps the first pair to the second pair), we can pick any of the inequalities
of type~\eqref{eq:fundamental:cone:ineq:2} to be fulfilled with
equality. Solving the edge enumeration problem for this $n-1$ dimensional cone
is simpler and using the automorphism group we can derive from this the
minimal pseudo-codewords of the cone of interest.

\subsection{Connection to Stopping Sets}
\label{sec:spectra:examples:connection:stopping:sets:1}

Before finishing this section that showed some pseudo-weight enumerators, let
us comment on stopping set weight enumerators for PG-based codes that were
investigated by Kashyap and Vardy~\cite{Kashyap:Vardy:03:1}. These stopping
set weight enumerators are tightly related to the BEC pseudo-weight
enumerators because of the following reasons~\cite{Koetter:Vontobel:03:1,
Vontobel:Koetter:05:1:subm, Feldman:Wainwright:Karger:05:1}: the support of
any pseudo-codeword is a stopping set whose size equals the BEC pseudo-weight
of the pseudo-codeword. Moreover, for every stopping set there is at least one
pseudo-codeword whose support equals that stopping set. However, some care
must be taken when comparing a BEC pseudo-weight enumerator and a stopping set
weight enumerator because the former is a sum over minimal pseudo-codewords
whereas the latter is a sum over minimal stopping sets. (See also the
discussion in Sec.~\ref{sec:effective:minimal:pseudo:codewords:1} on effective
minimal pseudo-codewords for the BEC.)

\section{Bounds on the AWGNC Pseudo-Weight}
\label{sec:bounds:awgnc:pseudo:weight:1}

In this section we will give some bounds on the AWGNC pseudo-weight. First we
discuss some general bounds on the pseudo-weight of arbitrary vectors in
$\Rp^n$; note that these bounds will depend only on the type of the
vector. Then we discuss some bounds for certain selected pseudo-codewords that
appear in the fundamental cone of PG-based codes. For easier notation, if no
confusion arises, we will use the shorter $\wps(\vomega)$ instead of
$\wpsAWGNC(\vomega)$.

\begin{Definition}
  \label{def:type:of:vector:1}

  Let $\vomega \in \Rp^n$ and let $\type_{\ell} \defeq \type_{\ell}(\vomega)$
  be the number of components of the vector $\vomega$ that are equal to
  $\ell$, where $\ell \in \Rp$. Then, we call $\vtype \defeq \vtype(\vomega) =
  (t_{\ell}(\vomega))_{\ell \in \Rp}$ the type of $\vomega$. (Note that in the
  following we do not assume that $\ell$ is a non-negative integer, only that
  it is a non-negative real number.)
\end{Definition}

It follows from this definition that only finitely many $\type_{\ell}$'s are
non-zero and that $\sum_{\ell} \type_{\ell} = |\set{I}|=n$ for any $\vomega \in
\Rp^n$. Moreover, because $|\supp(\vomega)|
= \sum_{\ell > 0} \type_{\ell}$, $\onenorm{\vomega} = \sum_{\ell} \ell
\type_{\ell}$, and $\twonorm{\vomega}^2 = \sum_{\ell} \ell^2 \type_{\ell}$,
we have
\begin{align*}
  \wpsAWGNC(\vomega)
    &= \frac{\left(\sum_{\ell} \ell \type_{\ell} \right)^2}
            {\sum_{\ell} \ell^2 \type_{\ell}}, {\rm ~~and~~}
  \wpsBEC(\vomega)
    = \sum_{\ell > 0} \type_{\ell}.
\end{align*}
If $\tilde \vomega = \alpha \cdot \vomega$ for some $\alpha \in \Rpp$ then its
type $\tilde \vtype \defeq \vtype(\tilde \vomega)$ is such that $\tilde
\type_{\alpha \ell} = \type_{\ell}$ for all $\ell$.

\begin{Lemma}
  \label{lemma:awgnc:pseudo:weight:lower:bound:1:1}

  Let $\vomega \in \Rp^n$ and let $\eta \neq 0$ be some arbitrary real
  number. Then
  \begin{align*}
    \wps(\vomega)
      &\geq
         \frac{2 \eta \onenorm{\vomega}
               -
               \twonorm{\vomega}^2}
              {\eta^2}
       = \frac{\sum\limits_{i=1}^{n} \omega_i (2 \eta - \omega_i)}
              {\eta^2}
  \end{align*}
  with equality if and only if $\vomega = \vect{0}$ or $\eta =
  \twonorm{\vomega}^2 / \onenorm{\vomega}$.
\end{Lemma}

\begin{Proof}
If $\vomega = \vect{0}$ then the statement is certainly true, so let us
assume that $\vomega \neq \vect{0}$. The square of any real number is
non-negative, therefore $\left(
   \eta \onenorm{\vomega} - \twonorm{\vomega}^2
 \right)^2
 \geq 0,$
with equality if and only if $\eta = \twonorm{\vomega}^2 /
\onenorm{\vomega}$, which, after rearranging, gives $\eta^2 \onenorm{\vomega}^2
     \geq
        2 \eta \onenorm{\vomega} \twonorm{\vomega}^2
        -
        \twonorm{\vomega}^4.$
Finally, dividing by $\eta^2 \twonorm{\vomega}^2$ and using the definition
of $\wps(\vomega)$, we obtain the desired result.

\end{Proof}

\begin{Corollary}
  \label{cor:awgnc:pseudo:weight:lower:bound:1:2}

  Let $\vomega \in \Rp^n$, let $\vtype \defeq \vtype(\vomega)$ be the type of
  $\vomega$, and let $\eta \neq 0$ be some arbitrary real number. Then
  \begin{align*}
    \wps(\vomega)
      &\geq \sum_{\ell}
              \beta_{\ell} \type_{\ell}
    \quad\quad
    \text{ with }
    \quad\quad
    \beta_{\ell}
       = \frac{\ell (2\eta - \ell)}
              {\eta^2}
       = 1
         -
         \left(
           1
           -
           \frac{\ell}{\eta}
         \right)^2.
  \end{align*}
\end{Corollary}

\begin{Proof}
  The result follows immediately from
  Lemma~\ref{lemma:awgnc:pseudo:weight:lower:bound:1:1}.
\end{Proof}

\begin{Corollary}
  \label{cor:awgnc:pseudo:weight:lower:bound:1:3}

  Let $\vomega \in \Rp^n$ and let $\vtype \defeq \vtype(\vomega)$. Moreover,
  let $r$ be the ratio of the largest positive $\ell$ such that $\type_{\ell}$
  is non-zero and the smallest positive $\ell$ such that $\type_{\ell}$ is
  non-zero. Then we have the lower bound
  \begin{align*}
    \wps(\vomega)
      &\geq
         \frac{4r}
              {(r+1)^2}
         \cdot
         |\supp(\vomega)|.
  \end{align*}
  This bound was also obtained by Wauer~\cite{Wauer:05:1} using a different
  derivation.
\end{Corollary}

\begin{Proof}
Let $m$ be the largest positive $\ell$ such that $\type_{\ell}$ is non-zero
and let $m'$ be the smallest positive $\ell$ such that $\type_{\ell}$ is
non-zero. These definitions obviously yield $r = m/m'$. Consider
Cor.~\ref{cor:awgnc:pseudo:weight:lower:bound:1:2} with $\eta =
\frac{m+m'}{2}$. We obtain $\wps(\vomega) \geq \sum_{\ell} \beta_{\ell}
\type_{\ell}$ $(\text{a})$ with $\beta_{\ell} = 4 \ell 
\frac{m + m' - \ell}{(m + m')^2} = 1 - (1 - \frac{2 \ell}{m + m'})^2$. 
We observe that $\beta_{m'} = \beta_m = \frac{4 m m'}{(m + m')^2} = \frac{4
r}{(r+1)^2}$. Since $\beta_{\ell}$ is strictly concave in ${\ell}$ we must
have $\beta_{\ell} > \beta_{m'} = \beta_{m} = \frac{4 r}{(r + 1)^2}$ for all
$m' < \ell < m$. It follows that 
\begin{align*}
  \wps(\vomega)
    &=
       \sum_{m' \leq \ell \leq m}
         \beta_{\ell} \type_{\ell}
     \geq \sum_{m' \leq \ell \leq m}
         \frac{4r}
            {(r+1)^2}
         \type_{\ell}
     = \frac{4r}
            {(r+1)^2}
         \sum_{m' \leq \ell \leq m}
           \type_{\ell}
     = \frac{4r}
            {(r+1)^2}
       \cdot
       |\supp(\vomega)|.
\end{align*}

\end{Proof}

\mbox{}

Under the same assumptions as in
Cor.~\ref{cor:awgnc:pseudo:weight:lower:bound:1:3}, Kelley and
Sridhara~\cite{Kelley:Sridhara:05:1:subm} proved that $\wps(\vomega) \geq
\frac{2 r^2}{(1+r^2)(r-1) + 2r} |\supp(\vomega)|$. Note that for $r =
1$ and $r = 2$ the bound in Cor.~\ref{cor:awgnc:pseudo:weight:lower:bound:1:3}
equals this bound and that for integers $r$ larger than $2$ the bound in
Cor.~\ref{cor:awgnc:pseudo:weight:lower:bound:1:3} is larger than this
bound. (Note that it follows from some simple considerations that for a
minimal pseudo-codeword $\vomega$ the ratio $r$ is $1$ or at least $2$.)

In the following subsections we will give bounds on the AWGNC pseudo-weight of
the class of minimal pseudo-codewords having the property that their
components are all small non-negative integers. Based on the examples in
Sec.~\ref{sec:spectra:examples:1} we formed the belief that for the Tanner
graph families under study that the non-codeword minimal pseudo-codewords with
small AWGNC pseudo-weight are from this class.

\subsection{Pseudo-Codewords with Zeros and Ones}

We start our analysis with pseudo-codewords of smallest possible entries,
i.e.~pseudo-codewords with zeros and ones. The following theorem gives a
lower bound on their weight.

\begin{Theorem}
  \label{theorem:awgnc:pseudo:weight:lower:bound:5:1}

  Let $\matr{H} \defeq \matr{H}_{\PGq}$ and let $\vomega \in \fch{K}{H}$ be a
  (not necessarily minimal) pseudo-codeword of type $\vtype$ with $t_0$
  non-negative, $\type_1$ positive, and $t_{\ell} = 0$ otherwise. If $\vomega$
  is not a codeword in $\code{C}_{\PGq}$ then
  \begin{align*}
    \wps(\vomega)
      &\geq
         \left\lceil
           \frac{q}{2} 
           + 
           1 
           + 
           \frac{1}{2}
           \sqrt{q^2 + 16q + 16}
         \right\rceil\geq q+4.
  \end{align*}
 
\end{Theorem}

\begin{Proof}
  See Sec.~\ref{sec:proof:theorem:awgnc:pseudo:weight:lower:bound:5:1}.
\end{Proof}

Note that the above bound yields, for $q = 2$ and $q = 4$, $\wps(\vomega) \geq
q+4$, and, for $q \geq 8$, $\wps(\vomega)\geq q + 5,$ respectively.

\begin{Remark}
  An example of a non-codeword pseudo-codeword with only zeros and ones as
  discussed in Th.~\ref{theorem:awgnc:pseudo:weight:lower:bound:5:1} was
  presented at the end of Sec.~\ref{sec:spectra:examples:pg:2:2:1}. Note,
  however, that this example was for $\matr{H}'_{\PG{2}{2}}$
  in~\eqref{eq:pg:2:2:modified:pcm:1} and not for $\matr{H}_{\PG{2}{2}}$
  in~\eqref{eq:pg:2:2:pcm:1}.
\end{Remark}

Observations for small $\PGq$-based codes suggest the following conjecture.

\begin{Conjecture}
  Let $\matr{H} \defeq \matr{H}_{\PGq}$ and let $\vomega \in \fch{K}{H}$ be a
  minimal pseudo-codeword of type $\vtype$ with both $t_0$ non-negative,
  $\type_1$ positive, and $t_{\ell} = 0$ otherwise. The vector $\vomega$ is
  then a minimal codeword.
\end{Conjecture}

The following theorem generalizes part 3 of
Lemma~\ref{lemma:binary:code:minimal:codeword:properties:1} (regarding
properties of minimal codewords) to pseudo-codewords with components equal
only to zero and one.

\begin{Lemma}
  Let $\code{C}$ be an $[n,k]$ binary linear code represented by a
  parity-check matrix $\matr{H}$. (Note that we do not assume that $\code{C} =
  \code{C}_{\PGq}$ or $\code{C} = \code{C}_{\EGq}$.) Let $\vomega \in
  \fch{K}{H}$ be a minimal pseudo-codeword of type $\vtype$ with $t_0$
  non-negative, $\type_1$ positive, and $t_{\ell} = 0$ otherwise.  Then
  $$\wps(\vomega)=\wH(\vomega)\leq n-k+1.$$
\end{Lemma}

\begin{Proof}
  Without loss of generality we can assume that the pseudo-codeword indices
  have been reordered such that the first $n_1 \defeq \wH(\vomega)$ components
  of $\vomega$ are equal to one and such that the remaining $n_2 \defeq n -
  \wH(\vomega)$ components of $\vomega$ are equal to zero.

  Let $\matr{K}_1 \vomega \geq \vect{0}$ be the collection of inequalities of
  type~\eqref{eq:fundamental:cone:ineq:1} that a pseudo-codeword must fulfill
  and let $\matr{K}_2 \vomega \geq \vect{0}$ be the collection of inequalities
  of type~\eqref{eq:fundamental:cone:ineq:2} that a pseudo-codeword must
  fulfill. Then there exists a full-rank $(n_1-1) \times n$-submatrix
  $\matr{A}_1 \defeq (\matr{A}_{11} \ | \ \matr{A}_{12})$ of $\matr{K}_1$ and
  a full-rank $n_2 \times n$-submatrix of $\matr{A}_2 \defeq (\matr{A}_{21} \
  | \ \matr{A}_{22})$ of $\matr{K}_2$ such that $\matr{A}_1 \vomega =
  \vect{0}$ and $\matr{A}_2 \vomega = \vect{0}$, such that $\matr{A}_{21} =
  \matr{0}$ and $\matr{A}_{22} = \matr{I}_{n_2}$, and such that
  $\rank_{\R}(\matr{A}) = n-1$, where $\matr{A} \defeq \bigl(
  \begin{smallmatrix} \matr{A}_1 \\ \matr{A}_2 \end{smallmatrix} \bigr) =
  \bigl( \begin{smallmatrix} \matr{A}_{11} & \matr{A}_{12} \\ \matr{A}_{21} &
  \matr{A}_{22} \end{smallmatrix} \bigr) = \bigl( \begin{smallmatrix}
  \matr{A}_{11} & \matr{A}_{12} \\ \matr{0} & \matr{I}_{n_2} \end{smallmatrix}
  \bigr)$.  Applying elementary row operations to the matrix $\matr{A} $ we
  obtain the matrix $\matr{\tilde A} = \bigl( \begin{smallmatrix}
  \matr{A}_{11} & \matr{0} \\ \matr{0} & \matr{I}_{n_2} \end{smallmatrix}
  \bigr)$. Because all non-zero entries of $\vomega$ are equal to one and
  because $\vomega$ is assumed to be a minimal pseudo-codeword, it turns out
  that all rows of $\matr{A}_{11}$ must contain exactly two non-zero elements,
  one equal to $+1$ and one equal to $-1$.

  These facts can be used as follows. First, we will show that
  $\rank_{\GF{2}}(\matr{A}_{11}) = n_1 - 1$. Secondly, we will show that
  $\rank_{\GF{2}}(\matr{A}_{11}) \leq n - k$. Finally, combining these results
  we will obtain the desired statement that $\wps(\vomega) = \wH(\vomega) =
  n_1 \leq n - k + 1$.
  
  So, let us show that $\rank_{\GF{2}}(\matr{A}_{11}) = n_1 - 1$, i.e.~that
  $\rank_{\GF{2}}(\matr{A}_{11}) = \rank_{\R}(\matr{A}_{11})$. Indeed, using
  the the special row structure of $\matr{A}_{11}$, it can be verified that
  the only vector in the (right-hand side) kernel of $\matr{A}_{11} \
  (\text{mod $2$})$ is the all-ones vector over $\GF{2}$ of length $n_1$.

  Secondly, let us show that $\rank_{\GF{2}}(\matr{A}_{11}) \leq n -
  k$. Indeed, we observe that every row in $\matr{A}_1 \ (\text{mod $2$})$
  corresponds to a row in $\matr{H}$. This implies that
  $\rank_{\GF{2}}(\matr{A}_{11}) \leq \rank_{\GF{2}}(\matr{A}_1) \leq
  \rank_{\GF{2}}(\matr{H}) \leq n - k$.
\end{Proof}

\begin{Remark}
  A crucial element in the above proof was the fact that all rows of
  $\matr{A}_{11}$ contain exactly two non-zero entries, one equal to $+1$ and
  one equal to $-1$. For minimal pseudo-codewords where not all non-zero
  entries are equal, this is not the case anymore and therefore we cannot use
  the above proof to generalize the lemma statement to other types of minimal
  pseudo-codewords.
\end{Remark}

\subsection{Pseudo-Codewords with Zeros, Ones, and Twos}

\begin{Theorem}
  \label{theorem:awgnc:pseudo:weight:lower:bound:3:1}

  Let $\matr{H} \defeq \matr{H}_{\PGq}$ and let $\vomega \in \fch{K}{H}$ be of
  type $\vtype$ with both $t_0$ non-negative, $\type_1$ positive, $\type_2$
  positive, and $t_{\ell} = 0$ otherwise. Then
  \begin{align*}
    \wps(\vomega)
      &\geq \frac{32}{27} (q+2)
       \approx    1.185 (q+2).
  \end{align*}
\end{Theorem}

\begin{Proof} 
For any $i \in \set{I}$ we must have $\sum_{i' \in \set{I} \setminus \{ i
\}} \omega_{i'} \overset{(\text{a})}{=} \sum_{j \in \set{J}_i} \sum_{i' \in
\set{I}_j \setminus \{ i \}} \omega_{i'} \overset{(\text{b})}{\geq} \sum_{j \in
\set{J}_i} \omega_i = (q+1) \omega_i$, where at step $(\text{a})$ we used 
the fact that all variable nodes are at graph distance two from each other in
the Tanner graph associated to $\matr{H}$, and where at step $(\text{b})$ we
used the inequalities in~\eqref{eq:fundamental:cone:ineq:1}. Adding $\omega_i$
to both sides we obtain $\sum_{i' \in \set{I}} \omega_{i'} \geq (q+2)
\omega_i$. Now, fix an $i \in \set{I}$ for which $\omega_i = 2$ holds and
express $\sum_{i' \in \set{I}} \omega_{i'}$ in terms of $\vtype$: it must hold
that $\type_1 + 2\type_2 \geq 2 (q+2)$ $(\text{c})$.

In a second step, we construct a vector $\vomega'=(\omega'_1, \ldots,
\omega'_n) \in \R^n$ such that
\begin{align*}
  \omega'_i
    &\defeq
       \begin{cases}
         0 & \text{if $\omega_i = 0$} \\
         2 & \text{if $\omega_i = 1$} \\
         1 & \text{if $\omega_i = 2$}
       \end{cases}
       \quad
       \text{(for all $i \in \set{I}$).}
\end{align*}
It can easily be seen that $\vomega'$ lies also in
the fundamental cone, i.e.~$\vomega' \in \fch{K}{H}$, and that $\vomega'$
has type $\vtype'$ with $t'_1 = \type_2$ positive, $t'_2 = \type_1$
positive, and $t'_{\ell} = t_{\ell}$ otherwise. In other words, switching $0
\mapsto 0$, $1 \mapsto 2$, $2 \mapsto 1$ we obtain another
pseudo-codeword. Arguing as above, for any $i \in \set{I}$ we must have
$\sum_{i' \in \set{I}} \omega'_{i'} \geq (q+2) \omega'_i$. Now, fix an $i
\in \set{I}$ for which $\omega'_i = 2$ holds, and express $\sum_{i' \in
\set{I}} \omega_{i'}$ in terms of $\vtype'$: it must hold that $\type'_1 +
2\type'_2 \geq 2 (q+2)$, i.e.~that $\type_2 + 2\type_1 \geq 2 (q+2)$
$(\text{d})$.

Combining $(\text{c})$ and $(\text{d})$ we obtain $3 (\type_1 +
\type_2) \geq 4(q+2)$, i.e.~$|\supp(\vomega)| = \type_1 + \type_2 \geq
\frac{4}{3} (q+2)$.  Using Cor.~\ref{cor:awgnc:pseudo:weight:lower:bound:1:3}
we can conclude that $\wps(\vomega)
    \geq \frac{4 \cdot 2}{(2+1)^2} \cdot \frac{4}{3} (q+2)
     =    \frac{8}{9} \cdot \frac{4}{3} (q+2)
     =    \frac{32}{27} (q+2).$

\end{Proof}

\mbox{}

Using some stronger assumptions on the pseudo-codeword $\vomega$ we can obtain
a stronger lower bound, as is shown in the next theorem.

\begin{Theorem}
  \label{theorem:awgnc:pseudo:weight:lower:bound:2:1}

  Let $\matr{H} \defeq \matr{H}_{\PGq}$ and let $\vomega \in \fch{K}{H}$ be of
  type $\vtype$ with both $t_0$ non-negative, $\type_1 \geq q+2$, $\type_2$
  positive, and $t_{\ell} = 0$ otherwise.\footnote{See
  Rem.~\ref{remark:qplustwo:vs:one:1} for a comment on these conditions.}  Then
  \begin{align*}
    \wps(\vomega)
      &\geq \frac{4}{3} (q+2)
       \approx   1.333 (q+2).
  \end{align*}
\end{Theorem}

\begin{Proof} The start is similar to the beginning of the proof of
Th.~\ref{theorem:awgnc:pseudo:weight:lower:bound:3:1}. For any $i \in
\set{I}$ we must have $\sum_{i' \in \set{I}} \omega_{i'} \geq (q+2)
\omega_i$. Now, fix an $i \in \set{I}$ for which $\omega_i = 2$ holds, and
express $\sum_{i' \in \set{I}} \omega_{i'}$ in terms of $\vtype$: it must hold
that $\type_1 + 2\type_2 \geq 2 (q+2)$, or, equivalently, $\type_2 \geq q + 2
- \type_1/2$. For any $\eta \neq 0$ we obtain
\begin{align*}
  \wps(\vomega)
    &\overset{(\text{a})}{\geq}
       \frac{(2\eta - 1) \type_1 + (4\eta - 4) \type_2}
            {\eta^2}
     \overset{(\text{b})}{\geq}
       \frac{(2\eta - 1) \type_1 + (4\eta - 4) (q + 2 - \type_1/2)}
               {\eta^2} \\
    &=    \frac{\type_1 + (4\eta - 4) (q + 2)}
               {\eta^2},
\end{align*}
where at step $(\text{a})$ we used
Cor.~\ref{cor:awgnc:pseudo:weight:lower:bound:1:2} and at step $(\text{b})$ we
used the inequality on $\type_2$ that we just found above. Using the
assumption that $\type_1 \geq q+2$ from the theorem statement we get
$\wps(\vomega) \geq \frac{(4\eta - 3) (q + 2)} {\eta^2}$. The right-hand side
of this expression is maximized by $\eta^{*} = \frac{3}{2}$: inserting this
value yields the lower bound in the theorem statement.

\end{Proof}

\begin{Remark}
  \label{remark:qplustwo:vs:one:1}

  Let $\code{C}$ be the code defined by $\matr{H}$. If a pseudo-codeword is an
  unscaled pseudo-codeword~\cite{Koetter:Li:Vontobel:Walker:04:1,
  Vontobel:Koetter:05:1:subm} then it is equal (modulo $2$) to a codeword of
  $\code{C}$. Therefore, the number of odd components of an unscaled
  pseudo-codeword must either be zero or at least equal to the minimum Hamming
  weight of the code. So, if we actually know that $\vomega$ in
  Th.~\ref{theorem:awgnc:pseudo:weight:lower:bound:2:1} is an unscaled
  pseudo-codeword then the requirement $\type_1 \geq q+2$ in the theorem
  statement is equivalent to the requirement $\type_1 \geq 1$.
  
  Note also that Th.~\ref{theorem:awgnc:pseudo:weight:lower:bound:2:1} can be
  generalized to the setup where $\vomega \in \fch{K}{H}$ has type $\vtype$
  with $t_0$ non-negative, $t_m$ positive for some integer $m \geq 2$,
  $t_{\ell}$ non-negative for $1 \leq \ell \leq m-1$, $t_{\ell} = 0$ for $\ell
  \geq m+1$, and $\sum_{\text{odd } \ell} t_{\ell} \geq q+2$. Then
  $\wps(\vomega) \geq\frac{m^2}{m^2-m+1} (q+2)$.
\end{Remark}

\mbox{}

\begin{figure}
  \begin{center}
    \epsfig{file=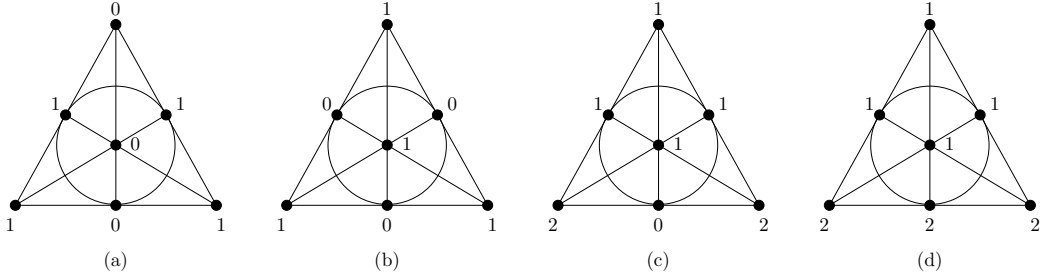, width=0.9\linewidth}
  \end{center}
  \caption{Codewords and pseudo-codewords used in
           Ex.~\ref{ex:adding:codewords:and:switching:1}}
  \label{fig:pg:2:2:pcw:1}
\end{figure}

\begin{Example}
  \label{ex:adding:codewords:and:switching:1}

  Let us exhibit some low-weight minimal pseudo-codewords that contain only
  zeros, ones, and twos. Consider first the case $q = 2$. The projective plane
  for $q = 2$ is shown in Fig.~\ref{fig:pg:2:2:pcw:1}~(a): it has $7$ points
  and $7$ lines and we consider the points to be variables and the lines to be
  checks. Fig.~\ref{fig:pg:2:2:pcw:1}~(a and b) shows two codewords of weight
  $q + 2 = 4$; note that their supports overlap in $\frac{q + 2}{2} = 2$
  positions. Adding these two codewords together yields the pseudo-codeword
  shown in Fig.~\ref{fig:pg:2:2:pcw:1}~(c). Switching the zero value into a
  two results in the pseudo-codeword in Fig.~\ref{fig:pg:2:2:pcw:1}~(d); it
  can be checked that this pseudo-codeword is actually a minimal
  pseudo-codeword. It has AWGNC pseudo-weight $6.25$, whereas the lower bounds
  in Ths.~\ref{theorem:awgnc:pseudo:weight:lower:bound:3:1}
  and~\ref{theorem:awgnc:pseudo:weight:lower:bound:2:1} are $4.74$ and $5.33$,
  respectively.

  Similarly, in the case of $q = 4$ it is possible to start with two codewords
  of weight $q+2 = 6$ whose supports overlap in $\frac{q + 2}{2} = 3$
  positions. After adding them and switching two zeros (that are
  specifically chosen and lie on the same line) into two twos, one gets a
  minimal pseudo-codeword of AWGNC pseudo-weight $9.85$, whereas the lower
  bounds in Ths.~\ref{theorem:awgnc:pseudo:weight:lower:bound:3:1}
  and~\ref{theorem:awgnc:pseudo:weight:lower:bound:2:1} are
  $7.11$ and $8.00$, respectively.

  In the case $q = 8$ it is possible to start with two codewords of weight
  $q+2 = 10$ whose supports overlap in $\frac{q + 2}{2} = 5$ positions. After
  adding them and switching three zeros (that are specifically chosen and form
  a triangle) into two twos, one gets a minimal pseudo-codeword of AWGNC
  pseudo-weight $16.10$, whereas the lower bound in
  Ths.~\ref{theorem:awgnc:pseudo:weight:lower:bound:3:1}
  and~\ref{theorem:awgnc:pseudo:weight:lower:bound:2:1} are $11.85$ and
  $13.33$, respectively.

  We conjecture that, with suitable generalizations, the above construction can
  be extended to larger $q$.
\end{Example}

\subsection{Pseudo-Codewords with Zeros, Ones, Twos, and Threes}

\begin{Theorem}
  \label{theorem:awgnc:pseudo:weight:lower:bound:4:1}

  Let $\matr{H} \defeq \matr{H}_{\PGq}$ and let $\vomega \in \fch{K}{H}$ be of
  type $\vtype$ with both $t_0$ non-negative, $\type_1$ positive, $\type_2$
  non-negative, $\type_3$ positive, and $t_{\ell} = 0$
  otherwise. We require that $\vomega$ is an unscaled pseudo-codeword.
  Then
  \begin{align*}
    \wps(\vomega)
      &\geq \frac{9}{8} \cdot (q+2)
       =    1.125 (q+2).
  \end{align*}
\end{Theorem}

\begin{Proof} The start is similar to the beginning of the proof of
Th.~\ref{theorem:awgnc:pseudo:weight:lower:bound:3:1}. For any $i \in
\set{I}$ we must have $\sum_{i' \in \set{I}} \omega_{i'} \geq (q+2)
\omega_i$. Now, fix an $i \in \set{I}$ for which $\omega_i = 3$ holds and
express $\sum_{i' \in \set{I}} \omega_{i'}$ in terms of $\vtype$: it must hold
that $\type_1 + 2\type_2 + 3\type_3 \geq 3 (q+2)$ $(\text{a})$.

In a second step, we construct a vector $\vomega'=(\omega'_1, \ldots,
\omega'_n) \in \R^n$ such that 
\begin{align*}
  \omega'_i
    &\defeq
       \begin{cases}
         3 & \text{if $\omega_i = 1$} \\
         2 & \text{if $\omega_i = 2$} \\
         1 & \text{if $\omega_i = 3$} \\
         0 & \text{otherwise}
       \end{cases}
       \quad
       \text{(for all $i \in \set{I}$).}
\end{align*}
It can be seen that $\vomega'$ lies also in the fundamental cone,
i.e.~$\vomega' \in \fch{K}{H}$,\footnote{Note that the inequality $3 + 3 \geq
1$ goes into the inequality $1 + 1 \geq 3$, which is wrong. However, we
assumed that $\vomega$ is an unscaled pseudo-codeword, which, among other
things, implies that the modulo-$2$ sum of the $\omega_i$'s that are involved
in a check is zero. Therefore, it cannot happen that the non-zero $\omega_i$'s
that are involved in a check have the values $3$, $3$, and $1$.} and that
$\vomega'$ has type $\vtype'$ with $\type'_1 = \type_3$ positive, $\type'_2 =
\type_2$ non-negative, $\type'_3 = \type_1$ positive, and 
$\type'_{\ell} = \type_{\ell}$ otherwise. Arguing as above, for any $i \in
\set{I}$ we must have $\sum_{i' \in \set{I}} \omega'_{i'} \geq (q+2) 
\omega'_i$. Now, fix an $i \in \set{I}$ for which $\omega'_i = 3$ holds 
and express $\sum_{i' \in \set{I}} \omega_{i'}$ in terms of $\vtype'$: it must
hold that $\type'_1 + 2\type'_2 + 3\type'_3 \geq 3 (q+2)$, i.e.~that $\type_3
+ 2\type_2 + 3\type_1 \geq 3 (q+2)$ $(\text{b})$.

Combining $(\text{a})$ and $(\text{b})$ we obtain $4 (\type_1 + \type_2 +
\type_3) \geq 6(q+2)$, i.e.~$|\supp(\vomega)| = \type_1 + \type_2 + \type_3
\geq \frac{3}{2} (q+2)$. Using 
Cor.~\ref{cor:awgnc:pseudo:weight:lower:bound:1:3} we can conclude that 
$\wps(\vomega)
    \geq \frac{4 \cdot 3}{(3+1)^2} \cdot \frac{3}{2} (q+2)
     =    \frac{3}{4} \cdot \frac{3}{2} (q+2)
     =    \frac{9}{8} (q+2).$

\end{Proof}

\section{Effective Minimal Pseudo-Codewords}
\label{sec:effective:minimal:pseudo:codewords:1}

The BSC can be seen as a binary-input AWGNC where the values at the output are
quantized to $+1$ or $-1$. It follows that the components of the
log-likelihood vector $\vlambda$ can take on only two values, namely $+L$ and
$-L$, where $L$ is a positive constant that depends on the bit flipping
probability of the BSC. Because of this quantization, there are certain things
that happen for the BSC that cannot happen for the AWGNC. Similarly, there are
things that happen for the BEC that cannot happen for the AWGNC or the BSC. By
introducing the so-called effectiveness of minimal pseudo-codewords, this
section discusses some of these special behaviors.

\begin{Definition}
  Fix a memoryless binary-input channel and let $\setL^{(n)} \subseteq (\R
  \cup \{ \pm \infty \})^n$ be the set of all possible log-likelihood ratio
  vectors upon sending the all-zero codeword.\footnote{For the AWGNC we have
  $\setLAWGNC^{(n)} = \R^n$, for the BSC we have $\setLBSC^{(n)} = \{ \pm L
  \}^n$ for some $L \in \Rpp$, and for the BEC we have $\setLBEC^{(n)} = \{ 0,
  +\infty \}^n$. Please note that there was a slight mistake
  in~\cite[Def.~10]{Vontobel:Smarandache:05:1},
  i.e.~in~\cite[Def.~10]{Vontobel:Smarandache:05:1} we forgot to require that
  the all-zeros codeword was sent. Nevertheless, compared
  to~\cite[Def.~10]{Vontobel:Smarandache:05:1}, the sets $\setLAWGNC^{(n)}$
  and $\setLBSC^{(n)}$ remain unchanged, whereas for the BEC we have $\{ 0,
  +\infty \}^n$ instead of $\{ -\infty, 0, +\infty \}^n$. Moreover, we assume
  that $L \in \Rpp$ and not $L \in \Rp$ in order to avoid the uninteresting
  case $L = 0$ stemming from a BSC with cross-over probability $1/2$.}
  Moreover, let us fix a parity-check matrix $\matr{H}$ and let
  $\Mps(\fch{K}{H})$ be the set of minimal pseudo-codewords. A minimal
  pseudo-codeword $\vomega \in \Mps(\fch{K}{H})$ is called \emph{effective of
  the first kind} for that particular channel if there exists a $\vlambda \in
  \setL^{(n)}$ such that $\langle \vomega, \vlambda \rangle < 0$ and $\langle
  \vomega', \vlambda \rangle \geq 0$ for all $\vomega' \in \Mps(\fch{K}{H})
  \setminus \{ \vomega \}$. A minimal pseudo-codeword $\vomega \in
  \Mps(\fch{K}{H})$ is called \emph{effective of the second kind} for that
  particular channel if there exists a $\vlambda \in \setL^{(n)}$ such that
  $\langle \vomega, \vlambda \rangle \leq 0$ and $\langle \vomega', \vlambda
  \rangle \geq 0$ for all $\vomega' \in \Mps(\fch{K}{H}) \setminus \{ \vomega
  \}$. (Obviously, a minimal pseudo-codeword that is effective of the first
  kind is also effective of the second kind.)
\end{Definition}

Let $\setL^{(n)}_{\vect{0}} \subseteq \setL^{(n)}$ be the set where LP
decoding decides in favor of the codeword $\vect{0}$. From the above
definition it follows that a minimal pseudo-codeword ``shapes'' the set
$\setL^{(n)}_{\vect{0}}$ if and only if it is an effective minimal
pseudo-codeword. More precisely, in the case where a minimal pseudo-codeword
$\vomega$ is effective of the first kind then there exists at least one
$\vlambda \in \setL^{(n)}$ where $\vomega$ wins against all other minimal
pseudo-codewords (and the zero codeword). Moreover, in the case where
$\vomega$ is effective of the second kind we are guaranteed that there is at
least one $\vlambda \in \setL^{(n)}$ were $\vomega$ is involved in a tie; if
and how often $\vomega$ wins against all other minimal pseudo-codewords (and
the zero codeword) depends on how ties are resolved.\footnote{The fact that a
minimal pseudo-codeword is effective of the second kind does of course not
exclude the possibility that there are also $\vlambda \in \setL^{(n)}$ were
$\vomega$ wins (unconditionally) against all other minimal pseudo-codewords
(and the zero codeword).}

\begin{Theorem}
  For the binary-input AWGNC and any parity-check matrix $\matr{H}$ all
  minimal pseudo-codewords of $\fch{K}{H}$ are effective of the first kind.
\end{Theorem}

\begin{Proof}
  This follows from basic cone properties
  (cf.~e.g.~\cite{Boyd:Vandenberghe:04:1}).
\end{Proof}

\mbox{}

We now turn to the BSC. As the following observations show, for this channel
not all minimal pseudo-codewords need to be effective of the first or of the
second kind.

\begin{Theorem}
  \label{theorem:bsc:correction:capabilities:1}

  Consider data transmission over a BSC using the code defined by $\matr{H}
  \defeq \matr{H}_{\PGq}$. Then LP decoding can correct any pattern of
  $\frac{q}{2}$ bit flips and no pattern of more than $q$ bit flips.
\end{Theorem}

\begin{Proof}
\label{sec:proof:theorem:bsc:correction:capabilities:1}
Because $\wpsBSCmin(\matr{H}) = q + 2$, the BSC pseudo-weight of any
pseudo-codeword in $\fch{K}{H}$ is at least $q + 2$. Therefore LP decoding
can correct at least $\lfloor \frac{q + 2 - 1}{2} \rfloor = \frac{q}{2}$ bit
flips.

Let us now show that LP decoding can correct at most $q$ bit flips. Remember
that a necessary condition for LP decoding to decode a received
log-likelihood vector $\vlambda$ to the zero codeword is that $\langle
\vomega, \vlambda \rangle \geq 0$ for all $\vomega \in
\fch{K}{H}$.\footnote{Note that this is usually not a sufficient condition
for correct decoding, e.g.~in the case where ties are resolved randomly.}
Assume that we are transmitting the zero codeword and that $e$ bit flips
happened. Hence $e$ components of $\vlambda$ are equal to $-L$ and $n-e$
components of $\vlambda$ are equal $+L$. It can easily be checked that the
following $\vomega$ is in $\fch{K}{H}$: let $\omega_i \defeq 1$ if $\lambda_i
= -L$ and $\omega_i \defeq 1/q$ otherwise. For this $\vomega$, the condition
$\langle \vomega, \vlambda \rangle \geq 0$ translates into $e \cdot (-L) +
(n-e) \cdot (1/q) \cdot (+L) \geq 0$, i.e.~$e \leq \frac{n}{q+1} = \frac{q^2 +
q + 1}{q+1} = q + \frac{1}{q+1}$. Because $e$ must be an integer this
inequality turns into the inequality $e \leq
\bigl\lfloor q + \frac{1}{q+1} \bigr\rfloor = q$.
\end{Proof}

Observe that the way we constructed the pseudo-codeword $\vomega$ in the proof
of Th.~\ref{theorem:bsc:correction:capabilities:1} can be seen as a
generalization of the so-called canonical
completion~\cite{Koetter:Vontobel:03:1, Vontobel:Koetter:05:1:subm}, however
instead of assigning values according to the graph distance with respect to a
single node, we assign values according to the graph distance with respect to
the set of nodes where $\lambda_i$ is negative. (Note that the Tanner graph of
$\matr{H} = \matr{H}_{\PGq}$ has a special property: all variable nodes are at
graph distance $2$ from each other.) Such a generalization of the canonical
completion was also used by Haley and Grant~\cite{Haley:Grant:05:1} for the
analysis of their codes.

\begin{Corollary}
  \label{cor:bounds:on:wpsBSC:for:effective:pcw:1}

  Consider the code defined by $\matr{H} \defeq \matr{H}_{\PGq}$. For the BSC,
  a necessary condition for a minimal pseudo-codeword $\vomega$ of
  $\fch{K}{H}$ to be effective of the second kind is that $q + 2 \leq
  \wpsBSC(\vomega) \leq 2q + 2$.
\end{Corollary}

\begin{Proof}
  See Sec.~\ref{sec:proof:cor:bounds:on:wpsBSC:for:effective:pcw:1}.
\end{Proof}

\mbox{}

For $q = 4$ it turns out that $\fc{K}(\matr{H}_{\PG{2}{4}})$ has minimal
pseudo-codewords with BSC pseudo-weight equal to $12$. (These minimal
pseudo-codewords have type $\vtype$ with $\type_2 = 1$, $\type_1 = 12$,
$\type_0 = 8$, and $t_{\ell} = 0$ otherwise.)
Cor.~\ref{cor:bounds:on:wpsBSC:for:effective:pcw:1} clearly shows that these
cannot be effective of the second kind for the BSC, since, for $q = 4$, any
effective minimal pseudo-codeword of the second kind must fulfill $6 \leq
\wpsBSC(\vomega) \leq 10$.

Judging from Fig.~\ref{fig:pg:code:q:4:min:pseudo:codewords:histogram:1} it
also seems --- as far as AWGNC and BSC pseudo-weight are comparable --- that
soft information is quite helpful for the LP decoder when decoding the code
$\code{C}_{\PG{2}{4}}$ defined by $\matr{H}_{\PG{2}{4}}$.

We now turn to the BEC. Because $\langle \vomega, \vlambda \rangle \geq 0$ for
all minimal pseudo-codewords $\vomega$, no minimal pseudo-codeword can be
effective of the first kind.

\begin{Theorem}
  \label{theorem:bec:correction:capabilities:1}

  Consider data transmission over a BEC using the code defined by
  $\matr{H}$. Let $\vomega$ be a minimal pseudo-codeword such that there is an
  unscaled pseudo-codeword associated to $\vomega$ with at least one odd
  component. Then there exists a non-zero codeword $\vc$ such that whenever
  $\langle \vomega, \vlambda \rangle = 0$ for some $\vlambda \in
  \setLBEC^{(n)}$ then also $\langle \vc, \vlambda \rangle = 0$.
\end{Theorem}

\begin{Proof}
\label{sec:proof:theorem:bec:correction:capabilities:1}
Let $\vomega'$ be an unscaled pseudo-codeword that is a positive multiple of
$\vomega$. By assumption, we can assume that at least one component of
$\vomega'$ is an odd integer. It follows~\cite{Vontobel:Koetter:05:1:subm}
that $\vc \defeq \vomega' \text{ (mod $2$)}$ is a non-zero codeword. Let
$\vlambda \in \setLBEC^{(n)}$ be such that $\langle \vomega, \vlambda \rangle
= 0$.  Because $\vomega'$ is a positive multiple of $\vomega$ we must have
$\langle \vomega', \vlambda \rangle = 0$ and because $\supp(\vc) \subseteq
\supp(\vomega')$ we must have $\langle \vc, \vlambda \rangle = 0$.
\end{Proof}

\begin{Corollary}
  \label{cor:theorem:bec:correction:capabilities:1}

  Consider data transmission over a BEC using the code defined by
  $\matr{H}$. Under block-wise ML decoding we define a block error to be the
  event that there is a tie among at least two codewords. Similarly, under LP
  decoding we define a block error to be the event that there is a tie among
  at least two pseudo-codewords. If for all minimal pseudo-codewords there
  exists an associated pseudo-codeword with at least one odd component then
  the block error rate of block-wise ML decoding coincides with the block
  error rate of LP decoding. \ecorollary
\end{Corollary}

By listing all the minimal pseudo-codewords, it can be shown numerically that
the condition in Cor.~\ref{cor:theorem:bec:correction:capabilities:1} is
fulfilled for $\matr{H} \defeq \matr{H}_{\PGq}$ when $q = 2$ and $q = 4$. It
follows that for these two codes block-wise ML and LP decoding yield the same
block error rate (under the above definition of block error rate). This
corroborates the observations made in
\cite[Fig.~1]{Zhu:Chugg:05:1} for $q = 4$.

\section{The Structure of Minimal Pseudo-Codewords}
\label{sec:structure:of:minimal:pseudo-codewords:1}

\begin{figure}
  \begin{center}
    \epsfig{file=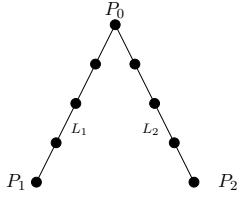, width=0.20\linewidth}
  \end{center}
  \caption{Part of $\PG{2}{4}$ discussed in
           Ex.~\ref{ex:structure:of:minimal:pcw:1}}
  \label{pg:2:4:pcw:1}
\end{figure}

In this section we discuss the geometry of minimal pseudo-codewords.
Minimum-weight codewords correspond to point-line configurations in the
projective plane that have been studied by several authors. Let us introduce
some notation and results from finite geometries,
cf.~e.g.~\cite{Hirschfeld:98:1}. A $k$-arc in $\PGq$ is a set of $k$ points no
three of which are collinear. A $k$-arc is complete if it is not contained in
a $(k+1)$-arc. The maximum number of points that a $k$-arc can have is denoted
by $m(2,q)$, and a $k$-arc with this number of points is called an oval (in
the case where $q$ is even this is sometimes also called a hyper-oval). One
can show that $m(2,q) = q + 2$ for $q$ even and $m(2,q) = q + 1$ for $q$
odd. One can make the following two interesting observations for the case $q$
even. Firstly, if two ovals have more than half their points in common, then
these two ovals coincide. Secondly, if a $q$-arc is contained in an oval then
the number of such ovals is one if $q > 2$ and two if $q = 2$.

It turns out that in the case $q$ even, the codewords with minimal weight are
$q+2$-arcs and therefore ovals. However, whereas the classification of ovals
for odd $q$ is simple (they all correspond to conics), the ovals for even $q$
are not classified that easily. For even $q$, one says that an oval is regular
if it comprises the points of a conic and its nucleus; one can show that for
$q = 2^s$, irregular ovals exist if and only if $s \geq 4$. It turns out that
the classification for irregular ovals is highly
non-trivial.\footnote{According to~\cite[Th.~8.35]{Hirschfeld:98:1}, there are
precisely two projectively distinct ovals in $\PG{2}{16}$, the so-called
regular oval $\set{D}(T^2)$ and the so-called oval $\set{O}_0 =
\set{D}(F_0)$. Moreover, according to~\cite[Th.~8.36]{Hirschfeld:98:1}, there
are precisely six projectively distinct ovals in $\PG{2}{32}$.} So, given that
even the classification of the codewords of minimal weight is difficult, it is
probably hopeless to obtain a complete classification of the minimal codewords
and minimal pseudo-codewords of codes defined by $\matr{H}_{\PGq}$, however it
is an interesting goal to try to understand as much as possible about the
structure of these codewords and pseudo-codewords.

In some recent papers, the structure of codewords of projective-plane-based
codes has been discussed by Kashyap and Vardy~\cite{Kashyap:Vardy:03:1} (that
paper talks also about stopping sets in Tanner graphs derived from projective
planes), by Justesen et al.~\cite{Justesen:Hoeholdt:Hjaltason:05:1,
Justesen:Hoeholdt:Hjaltason:05:2:subm} and by Laendner and
Milenkovic~\cite{Laendner:Milenkovic:05:1} (that paper also talks about
trapping sets in Tanner graphs derived from projective planes). Moreover, the
minimal-weight codewords of Euclidean-plane-based codes were discussed by
H{\o}holdt et al.~\cite{Hoeholdt:Justesen:Jonsson:05:1}; also here these
configurations are tightly related to ovals. However, because not all ovals
are regular, the classification is not that simple also for these
codes.\footnote{Note that the remark ``It is also known that an oval in $\EGq$
consists of a conic and a nucleus'' (which would imply that they are regular)
in~\cite[Sec.~2]{Hoeholdt:Justesen:Jonsson:05:1} is wrong in
general~\cite{Hoeholdt:05:1}.}

From now on, we will only consider projective planes $\PGq$ and $q$ will
always be even, i.e.~a power of two. Before we state our conjecture about the
structure of minimal pseudo-codewords, let us first look at an example.

\begin{Example}
  \label{ex:structure:of:minimal:pcw:1}
  
  Let $q=4$. We can find a minimal pseudo-codeword $\vomega$ whose type
  $\vtype$ is $\type_0 = 8$, $\type_1 = 8$, $\type_2 = 5$, and $\type_{\ell} =
  0$ otherwise. This pseudo-codeword can be obtained using a procedure similar
  to the one used in Ex.~\ref{ex:adding:codewords:and:switching:1}. Firstly,
  on has to add two vectors $\vx^{(1)}$ and $\vx^{(2)}$ of weight $6$ whose
  supports overlap in two positions. This yields a pseudo-codeword $\tilde
  \vomega$ of type $\tilde \vtype$ with $\tilde \type_0 = 11$, $\tilde \type_1
  = 8$, $\tilde \type_2 = 2$, and $\tilde \type_{\ell} = 0$ otherwise.
  Secondly, in order to obtain a minimal pseudo-codeword, one has to switch
  three zeros (that were appropriately chosen) into three twos.

  Let us analyze this procedure. Since a minimal pseudo-codeword corresponds
  to an edge of the fundamental cone, it is clear that the inequalities
  in~\eqref{eq:fundamental:cone:ineq:1} and~\eqref{eq:fundamental:cone:ineq:2}
  that are fulfilled with equality must form a system of linear equations of
  rank $21 - 1 = 20$. We start with two minimal codewords $\vx^{(1)}$ and
  $\vx^{(2)}$ that each yield a system of linear equations of rank $21 - 1 =
  20$. These two codewords have been chosen such that their sum $\tilde
  \vomega$ yields a system of linear equations of rank $21 - 2 = 19$.

  To find the three zeros that we have to switch, we proceed as follows. It
  turns out that in the projective plane $\PG{2}{4}$ there are two lines, say
  $L_1$ and $L_2$, such that all the entries of $\tilde \vomega$ that
  correspond to the points on these two lines are zero. Let $P_0$ be the
  intersection point of these two lines,
  cf.~Fig.~\ref{pg:2:4:pcw:1}. There exists a point $P_1$
  on $L_1$ and a point $P_2$ on $L_2$ such that modifying $\tilde \vomega$ by
  assigning them the same value $\alpha \geq 0$ yields a vector in the
  fundamental cone, as long as $\alpha$ is not too large. In fact, for $\alpha
  > 2$ the vector is outside the fundamental cone, and for $\alpha = 2$ it
  yields a vector that is a pseudo-codeword and that yields a system of
  equations of rank $21 - 1 = 20$, i.e.~it is a minimal
  pseudo-codeword.
\end{Example}

\begin{Conjecture}
  For the Tanner graph defined by $\matr{H}_{\PGq}$ every minimal
  pseudo-codeword is a sum of a few minimal pseudo-codewords with a change of
  a few low-value components such that they become the large components in the
  equations associated to the lines that pass through them.

  Hence, to find minimal pseudo-codewords, we have to take sums of two minimal
  pseudo-codewords that give rank $n-2$ (if possible; lower otherwise) and
  change few components that are ``not significant'' into a ``significant''
  ones. We call a component significant if it is the sum of the other
  components that belong to a line passing through the point, for most of such
  lines.
\end{Conjecture}

The following conjecture is based on our observations (for small $q$) that
among the minimal pseudo-codewords the ones with only zeros, ones, and twos
yield non-codeword minimal pseudo-codewords of smallest weight. A positive
answer to this conjecture would result in a much better understanding of the
minimal pseudo-codewords in general and of the AWGNC pseudo-weight spectrum
gap, in particular.

\begin{Conjecture}
  Let $\matr{H} \defeq \matr{H}_{\PGq}$ and consider the pseudo-codewords that
  have minimal AWGNC pseudo-weight among all minimal pseudo-codewords that are
  not multiples of minimal codewords. We conjecture that the type $\vtype$ of
  these pseudo-codewords is such that $\type_0$ is non-negative, $\type_1$ is
  positive, $\type_2$ is positive, and $\type_{\ell} = 0$ otherwise. (If this
  conjecture is not true, find the the smallest $\tilde \ell$ such that these
  pseudo-codewords have type $\vtype$ with $t_{\ell} \geq 0$ for $\ell \in \{
  0, 1, \ldots, \tilde \ell \}$ and $\type_{\ell} = 0$ otherwise.)
\end{Conjecture}

\section{Conclusions}
\label{sec:conclusions:1}

We have investigated the minimal pseudo-codewords of some codes whose Tanner
graphs are derived from projective and Euclidean planes and we have introduced
the notion of a pseudo-weight spectrum gap for a parity-check matrix, a
concept which is certainly worthwhile to be further explored. Although our
numerical results are for codes of modest length, to the best of our knowledge
this is the first study that tries to \emph{analytically} quantify the
behavior of $\PGq$- and $\EGq$-based binary linear codes under LP and
iterative decoding. Extending these results to somewhat longer codes has the
potential to explain many experimental observations made in the
past. Moreover, we have obtained a clearer picture about the structure of the
minimal pseudo-codewords of the Tanner graphs under investigation,
nevertheless more work is required to get a sufficiently tight
characterization of them.

We also introduced the notion of the effectiveness of a minimal
pseudo-codeword and we saw that in the case of non-AWGNC channels there are
minimal pseudo-codewords that are not effective (of the first kind or of the
second kind). Interestingly, in that context we were able to use the canonical
completion, a tool that so far has been very useful for characterizing
families of $(\wcol, \wrow)$-regular LDPC codes, with $\wcol,~\wrow$ bounded
when the block length goes to infinity, i.e.~code families where the Tanner
graph diameter grows with the block length. (This is in contrast to the
$\PGq$-based Tanner graphs which have girth $6$ and diameter $3$,
independently of $q$).

\appendix

\section{Proofs}
\label{sec:proofs:1}

We include here two of the larger proofs of this paper.

\subsection{Proof of Theorem~\ref{theorem:awgnc:pseudo:weight:lower:bound:5:1}}
\label{sec:proof:theorem:awgnc:pseudo:weight:lower:bound:5:1}

\begin{figure}
  \begin{center}
    \epsfig{file=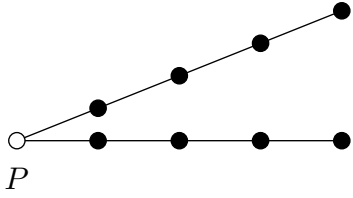,
            width=0.3\linewidth}
  \end{center}
  \caption{Part of $\PGq$ showing the relevant part of a codeword with Hamming
    weight $2q$. (Here for $q = 4$.)}
  \label{fig:proj:plane:weight:2q:codeword:1}
\end{figure}

\begin{figure}
  \begin{center}
    \epsfig{file=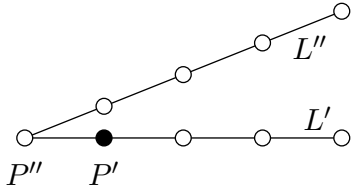,
            width=0.3\linewidth}
  \end{center}
  \caption{Part of $\PGq$ showing the points $P'$ and $P''$ and the lines $L'$
    and $L''$ that were used in proof of
    Th.~\ref{theorem:awgnc:pseudo:weight:lower:bound:5:1}. (Here for $q =
    4$.)}
  \label{fig:proj:plane:weight:find:configuration:1}
\end{figure}

Let $\vs^\tr \defeq \matr{H} \cdot \vomega^\tr \text{ (in $\R$)}$ and let
$\setJodd \defeq \{ j \in \set{J} \ | \ s_j \text{ is odd} \}$ be the set of
all rows of $\matr{H}$ where the corresponding entry in $\vs$ is an odd
integer.  If $\vomega$ were a codeword then all entries of $\vs$ would be
non-negative even integers. However, because $\vomega$ is assumed to be a
non-codeword, $|\setJodd| \geq 1$. Note moreover that because $\supp(\vomega)$
is a stopping set, we must have $s_j \neq 1$ for all $j \in \set{J}$.

The proof proceeds as follows. First we will show that for any non-codeword
$\vomega$ we must have $|\setJodd| \geq q+1$. Secondly, we will show that
this lower bound on $|\setJodd|$ implies the lower bound mentioned in the
theorem.

So, let us show that $|\setJodd| \geq q + 1$ for any non-codeword
$\vomega$. The proof is by contradiction, i.e.~if $\setJodd$ has $|\setJodd|
\leq q$, we will show that there exists a $j \in \setJodd$ such that the value
of $s_j$ must be an even integer. To that end, it turns out to be useful to
reverse the usual interpretation of the columns and rows of $\matr{H}$: the
columns will correspond to lines and the rows will correspond to points of
$\PGq$. With this, there is a one-to-one relationship between the points of
$\PGq$ and the entries of $\vs$.

Before we proceed, consider Fig.~\ref{fig:proj:plane:weight:2q:codeword:1}
that shows two lines in $\PGq$ and the points on them (here for $q =
4$). Letting $\vx \in \R^n$ be a vector where the entries corresponding to
black dots equal $1$ and the other entries equal $0$, one can easily verify
that $\vect{x} \cdot \matr{H} =
\vect{0} \text{ (mod $2$)}$, i.e.~$\vx$ is a codeword of the code with 
parity-check matrix $\matr{H}^\tr$.

Now, choose any $j \in \setJodd$, consider
Fig.~\ref{fig:proj:plane:weight:find:configuration:1}, and let $P'$ correspond
to the $j$-th row of $\matr{H}$. Because $|\setJodd| \leq q$, it is possible
to choose a line $L'$ through $P'$ such that all points on it (except for
$P'$) have an even $s_j$. On this line it is then possible to choose (again
because $|\setJodd| \leq q$) a $P''
\neq P'$ such that there is a line $L''$ through it such that all points on it
have an even $s_j$. Let $\vx \in \R^n$ be a vector where the non-zero entries
corresponding to the points on $L'$ and $L''$ (except for $P''$) equal $1$ and
the other entries equal $0$. Because of the considerations in the paragraph
above, it is clear that $\vect{x} \cdot \matr{H} = \vect{0} \text{ (mod
  $2$)}$, i.e.~all entries in $\vx \cdot \matr{H}$ are even integers. This
implies that $\vx \cdot \vs^\tr = \vx \cdot (\matr{H} \cdot \vomega^\tr) =
(\vx \cdot \matr{H}) \cdot \vomega^\tr$ must be an even integer. This is a
contradiction, because of the way we have chosen $P'$, $L'$, $P''$, and $L''$,
the inner product $\vx \cdot \vs^\tr$ must be an odd integer.

Let us now prove that the lower bound $|\setJodd| \geq q+1$ on $|\setJodd|$
implies the lower bound on $\wps(\vomega)$ mentioned in the theorem. Because
of the special properties of $\matr{H}$ we have $\matr{H}^\tr \matr{H} = q
\matr{I} + \matr{J}$, where $\matr{I}$ is the identity matrix of size 
$n \times n$ and where $\matr{J}$ is the all-ones matrix of size $n \times n$.
Then $\twonorm{\vs}^2 = \vs \vs^\tr = \vomega \matr{H}^\tr \matr{H}
\vomega^\tr = q \vomega \matr{I} \vomega^\tr + \vomega \matr{J} \vomega^\tr =
q \twonorm{\vomega}^2 + \onenorm{\vomega}^2 = q \type_1 + \type_1^2$. On the
other hand,
\begin{align*}
  \twonorm{\vs}^2
    &= \sum_{j \in \set{J}}
         s_j^2 
     = \sum_{j \in \set{J}} 
         \left(
           \sum_{i \in \set{I}_j} \omega_i
         \right)^2
     = \sum_{j \in \set{J}} 
         \sum_{i \in \set{I}_j} 
           \omega_i 
           \sum_{i' \in \set{I}_j} 
             \omega_{i'}.
\end{align*}
We would like to find a lower bound on $\twonorm{\vs}^2$. If $j \in \set{J}
\setminus \setJodd$ then we use $\sum_{i' \in \set{I}_j} \omega_{i'} \geq 2
\omega_i$, 
 for all $i$ in $\set{I}_j$
(which is implied by \eqref{eq:fundamental:cone:ineq:1}), otherwise
we use $\sum_{i' \in \set{I}_j} \omega_{i'} \geq 2 \omega_i + 1$ (which also
follows from \eqref{eq:fundamental:cone:ineq:1}, together with the observation
that $s_j = \sum_{i' \in \set{I}_j} \omega_{i'} \geq 3$):
\begin{align*}
  \twonorm{\vs}^2
    &= \sum_{j \in \set{J}} 
         \sum_{i \in \set{I}_j} 
           \omega_i 
           \sum_{i' \in \set{I}_j} 
               \omega_{i'}
     \geq
       2
       \sum_{j \in \set{J}} 
         \sum_{i \in \set{I}_j} 
           \omega_i^2
       +
       \sum_{j \in \setJodd} 
         \sum_{i \in \set{I}_j} 
           \omega_i
     = 2
       \sum_{i \in \set{I}} 
         \sum_{j \in \set{J}_i} 
           \omega_i^2
       +
       \sum_{j \in \setJodd} 
         \sum_{i \in \set{I}_j} 
           \omega_i \\
    &= 2 (q+1) \twonorm{\vomega}^2
       +
       \sum_{j \in \setJodd} 
         \sum_{i \in \set{I}_j} 
           \omega_i
     \geq
       2 (q+1) \twonorm{\vomega}^2
       +
       3 |\setJodd|
     = 2 (q+1) \type_1
       +
       3 |\setJodd|.
\end{align*}
Combining the above results we obtain $q \type_1 + \type_1^2 =
\twonorm{\vs}^2 \geq 2 (q+1) \type_1 + 3 |\setJodd|$, or, equivalently,
$\type_1^2 - (q+2) \type_1 - 3 |\setJodd| \geq 0$. It follows that $\type_1
\geq \frac{q}{2} + 1 + \frac{1}{2}\sqrt{(q+2)^2+12
|\setJodd|}$.\footnote{$\type_1 \leq \frac{q}{2} + 1 -
\frac{1}{2}\sqrt{(q+2)^2+12 |\setJodd|}$ is not possible because we know
that $\type_1 \geq q+2$.} Inserting the lower bound $|\setJodd| \geq q+1$, we
obtain
\begin{align*}
  \type_1
    &\geq
       \frac{q}{2} 
         + 
       1 
       + 
       \frac{1}{2}
       \sqrt{q^2 + 16q + 16}.
\end{align*}
Because $\wps(\vomega) = \type_1$ and because $\type_1$ is an integer, the
final result follows.

We conclude with two remarks:
\begin{itemize}

  \item More sophisticated considerations might lead to better lower bounds on
    $\wps(\vomega)$; however, note that $|\setJodd| \geq q+1$ is the best
    lower bound that can be given on the size of $\setJodd$ without additional
    information about the set. Namely, if $\setJodd$ happens to be equal to
    $\set{J}_i$ for some $i \in \set{I}$ then $|\setJodd| = q+1$ and no row in
    $\matr{H}$ corresponding to an entry in $\setJodd$ can be expressed as a
    linear combination of rows corresponding to entries in $\set{J} \setminus
    \setJodd$.
  \item The bounding techniques used in the second part of the proof were
    inspired by the bounding techniques that were used
    in~\cite{Vontobel:Koetter:04:1}, which in turn were generalizations
    of~\cite{Tanner:01:1}.

\end{itemize}

\subsection{Proof of 
            Corollary~\ref{cor:bounds:on:wpsBSC:for:effective:pcw:1}}
\label{sec:proof:cor:bounds:on:wpsBSC:for:effective:pcw:1}

Let $\vomega$ be a minimal pseudo-codeword with $\wpsBSC(\vomega) > 2q +
2$. The proof is by contradiction, i.e.~we will assume that $\vomega$ is
effective of the second kind and then we will show that for any $\vlambda \in
\setLBSC^{(n)}$ with $\langle \vomega, \vlambda \rangle \leq 0$ there exists 
a minimal pseudo-codeword $\vomega'' \neq \vomega$ such that $\langle
\vomega'', \vlambda \rangle < 0$.

Assume that we are transmitting the zero codeword and that bit flips happened
at positions $\set{E}$. Hence $|\set{E}|$ components of $\vlambda$ are equal
to $-L$ and $n-|\set{E}|$ components of $\vlambda$ are equal $+L$. Assume that
$\langle \vomega, \vlambda \rangle \leq 0$. From the definition of the BSC
pseudo-weight it follows that $|\set{E}| > q+1$, i.e.~$|\set{E}| \geq q+2$.

Choose a sub-pattern $\set{E}' \subset \set{E}$ of bit-flips with $|\set{E}'|
= q+1$ and define the corresponding $\vlambda'$. From the definition of the
BSC pseudo-weight it follows that $\langle \vomega,
\vlambda' \rangle > 0$ (otherwise $\wpsBSC(\vomega) \leq 2q+2$).

Similarly to the proof of Theorem~\ref{theorem:bsc:correction:capabilities:1},
we can construct a pseudo-codeword $\vomega'$ based on $\set{E}'$ such that
$\vomega'_i \defeq 1$ if $i \in \set{E}'$ and $\vomega'_i \defeq 1/q$
otherwise. This pseudo-codeword has the property that $\langle
\vomega', \vlambda' \rangle = |\set{E}'| \cdot (-L) + (n - |\set{E'}|) \cdot
(1/q) \cdot (+L) = - L < 0$. Let $\{ \vomega^{(\ell)} \}_{\ell}$ be the set of
all minimal pseudo-codewords. Then $\vomega' = \sum_{\ell}
\alpha_{\ell} \vomega^{(\ell)}$ for some choice of 
$\{ \alpha_{\ell} \}_{\ell}$ where all $\alpha_{\ell}$ are
non-negative. Therefore, there exists at least one minimal pseudo-codeword,
say $\vomega''$, such that $\langle \vomega'',
\vlambda' \rangle < 0$. (It is clear that $\vomega'' \neq \vomega$.)

Because $\vomega'' \geq \vect{0}$, it is easy to see that $\langle \vomega'',
\vlambda \rangle \leq \langle \vomega'', \vlambda' \rangle$. This implies 
that $\langle \vomega'', \vlambda \rangle < 0$, which is the promised
contradiction.

\bibliographystyle{ieeetr}

\bibliography{/home/vontobel/references/references}

\end{document}